\documentclass[12pt]{article}
\usepackage{ametsoc}
\usepackage{lineno}
\linenumbers

\usepackage{natbib}
\DeclareMathOperator*{\argmin}{argmin}

\newcommand{\bbP}{\mathbb P}
\newcommand{\bbE}{\mathbb E}
\newcommand{\cN}{{\mathcal N}}
\newcommand{\cC}{{\mathcal C}}

\newcommand{\cM}{{\mathcal M}}
\newcommand{\cH}{{\mathcal H}}

\newcommand{\cQ}{{\mathcal Q}}

\newcommand{\myabstract}{
Data assimilation leads naturally to a Bayesian formulation in which the posterior probability distribution of the system state, given all the observations on a time window of interest, plays a central conceptual role. The aim of this paper is to use this Bayesian posterior probability distribution as a gold standard against which to evaluate various commonly used data assimilation algorithms. 

A key aspect of geophysical data assimilation is the high dimensionality and limited predictability of the computational model.  We study the 2D Navier-Stokes equations in a periodic geometry, which has these features and yet is tractable for explicit and accurate computation of the posterior distribution by state-of-the-art statistical sampling techniques. The commonly used algorithms that we evaluate, as quantified by the relative error in reproducing moments of the posterior, are 4DVAR and a variety of sequential filtering approximations based on 3DVAR and on extended and ensemble Kalman filters.
  
The primary conclusions are that {\it under the assumption of a well-defined posterior probability distribution}: (i) with appropriate parameter choices, approximate filters can perform well in reproducing the mean of the desired probability distribution; (ii) however they do not perform as well in reproducing the covariance; (iii) the error is compounded by the need to modify the covariance, in order to induce stability.  Thus, filters can be a useful tool in predicting mean behavior, but should be viewed with caution as predictors of uncertainty. These conclusions are intrinsic to the algorithms when assumptions underlying them are not valid and will not change if the model complexity is increased.

}

\begin{document}

\title{\textbf{\large{Evaluating Data Assimilation Algorithms}}}

\author{\textsc{K.\ J.\ H.\ Law}
				\thanks{\textit{Corresponding author address:} 
Kody J. H. Law, Warwick Mathematics Institute, University of Warwick,
Coventry CV4 7AL, UK.
				\newline{E-mail: k.j.h.law@warwick.ac.uk}}\quad\textsc{and A.\ M.\ Stuart}\\
\textit{\footnotesize{Warwick Mathematics Institute, University of Warwick,
Coventry, UK}} }

\ifthenelse{\boolean{dc}}
{
\twocolumn[
\begin{@twocolumnfalse}
\amstitle

\begin{center}
\begin{minipage}{13.0cm}
\begin{abstract}
	\myabstract
	\newline
	\begin{center}
		\rule{38mm}{0.2mm}
	\end{center}
\end{abstract}
\end{minipage}
\end{center}
\end{@twocolumnfalse}
]
}
{
\amstitle
\begin{abstract}
\myabstract
\end{abstract}
\newpage
}


\section{Introduction}
\label{intro}

The positive impact of data assimilation schemes
on numerical weather prediction (NWP) is unquestionable.
Improvements in forecast skill over decades
reflect not only the increased resolution of the computational
model, but also the increasing volumes of data available,
and the increasing sophistication of algorithms to
incorporate this data. However, because of the huge
scale of the computational model, many of the algorithms
used for data assimilation employ 
approximations, based on both physical
insight and computational expediency, whose effect can be
hard to evaluate. The aim of this paper is to describe
a method of evaluating some important aspects of data
assimilation algorithms, by comparing them with a gold-standard:
the Bayesian posterior probability distribution on 
the system state given observations. In so doing we will
demonstrate that carefully chosen filters can perform
well in predicting mean behaviour, but that they
typically perform poorly when predicting uncertainty,
such as covariance information. 

In typical
operational conditions the observed data, model initial conditions,
and model equations are all subject to uncertainty. Thus we
take the perspective that the gold standard, which we
wish to reproduce as accurately as possible, is the
(Bayesian) posterior probability distribution of the system
state (possibly including parameters) given the observations. 
For practical weather forecasting scenarios this is not
computable. The two primary competing methodologies for 
data assimilation that {\em are} computable, and hence are
implemented in practice,
are {\em filters} \cite{kalnay2003atmospheric} and 
{\em variational methods} \cite{ben02}. We will compare
the (accurately computed, extremely expensive) Bayesian 
posterior distribution 
with the output of the (approximate, relatively cheap)
filters and variational methods used in practice. Our
underlying dynamical model is the 2D Navier-Stokes equations 
in a periodic setting.  This provides a high
dimensional dynamical system, which exhibits a range of
complex behaviours, yet which is sufficiently small that
the Bayesian posterior may be accurately computed by
state-of-the-art statistical sampling in an off-line
setting.

The idea behind filtering is to update the
posterior distribution of the system state sequentially
at each observation time. This may be performed exactly
for linear systems subject to Gaussian noise, and is
then known as the Kalman filter 
\cite{kalman1960new, harvey1991forecasting}. 
For nonlinear or non-Gaussian scenarios the particle
filter \cite{doucet2001sequential} 
may be used and provably approximates the desired 
probability distribution as the number of particles 
is increased \cite{bain2008fundamentals}. However 
in practice this method performs poorly in high
dimensional systems \cite{SBBA08} and, 
whilst there is considerable research activity
aimed at overcoming this degenertation 
\cite{van2010nonlinear, chorin2010implicit, bengtsson2003toward}, 
it cannot currently be viewed as a practical tool 
within the context of geophysical data assimilation. 
In order to circumvent problems associated
with the representation of high dimensional probability
distributions some form of Gaussian approximation is typically
used to create practical filters.
The oldest and simplest such option is to use a nonlinear
generalization of the mean update in the Kalman filter,
employing a constant 
prior covariance operator, obtained offline 
through knowledge coming from the underlying model and 
past observations \cite{lorenc1986analysis}; this methodology
is sometimes refered to as {\em 3DVAR}.
More sophisticated approximate Gaussian 
filters arise from either linearizing the dynamical model,
yielding the {\em extended Kalman filter} \cite{jazwinski1970stochastic},
or utilizing ensemble statistics, leading to the
{\em ensemble Kalman filter} 
\cite{evensen1994assimilation, evensen2003ensemble}.
Information about the underlying local (in time) Lyapunov
vectors, or bred vectors (see \cite{kalnay2003atmospheric}
for discussion) can be used to guide further approximations that
are made when implementing these methods in high dimensions. 
We will also be interested in the use of
{\em Fourier diagonal filters}, introduced in
\cite{harlim2008filtering,majda2010mathematical}, 
which approximate  
the dynamical model by a statistically equivalent linear 
dynamical system in a manner which enables 
the covariance operator to  
be mapped forward in closed form; in steady state the version we
employ here reduces to a particular choice of 3DVAR, based
on climatological statistics.  
An overview of particle filtering for geophysical systems
may be found in \cite{VL09} and 
a quick introduction to sequential filtering may be found in
\cite{arulampalam2002tutorial}.

Whilst 
filtering updates the system state sequentially each time
  when a new observation becomes available
variational methods attempt to incorporate data
which is distributed over an entire time-interval. This may be
viewed as an optimization problem where the objective function
is to choose the initial state, and possibly forcing to the
physical model, in order to best match the data over
the specified time-window. 
As such it may be viewed as a PDE-constrained optimization 
problem \cite{hinze2008optimization}, and more
generally as a particular class of regularized
inverse problem 
\cite{vogel2002computational, tarantola2005inverse,
banks1989estimation}. This approach is referred to as
4DVAR in the geophysical literature, when the optimization
is performed over just the initial state of the system
\cite{talagrand1987variational,courtier1987variational}
and as weak constraint 4DVAR when optimization is also
over forcing to the system \cite{Z97}.

From a Bayesian perspective, the solution to an inverse problem
is statistical, rather than deterministic, and is hence
significantly more challenging: regularization is imposed
through viewing the unknown as a random variable, and the
aim is to find the
posterior probability distribution on the state of
the system on a given time window, given the observations 
on that time window. 
With the current and growing capacity of computers it is 
becoming relevant and tractable to begin to explore such 
approaches to inverse problems in differential
equations \cite{kaipio2005statistical},
even though it is currently infeasible to do so for NWP.
There has, however, been some limited study of
the Bayesian approach to inverse problems in fluid
mechanics using path integral formulations in
continuous time as introduced in \cite{HSV07a}; see 
\cite{ApteJ07, apte2008data, quinn2010state, CDS11}
for further developments. 
We will build on the algorithmic experience 
contained in these papers here.  
For a recent overview of Bayesian methodology for
inverse problems in differential equations, see
\cite{stuart2010inverse}, and for the Bayesian
formulation of a variety of inverse problems
arising in fluid mechanics see \cite{cotter2009bayesian}.
The key take home message of this body of work on
Bayesian inverse problems is that it is often
possible to compute the posterior distribution of
state given noisy data with high degree of accuracy,
albeit at great expense: the methodology could not
be used online as a practical algorithm, but provides
us with a gold-standard against which we can evaluate
on-line approximate methods used in practice.

There are several useful connections to make between the
Bayesian posterior distribution, filtering methods
and variational methods all of which serve to highlight the
fact that they are all {\em attempting} to represent related
quantities. 
The first observation is that, in the
linear Gaussian setting, if backward filtering is
implemented on a given time window (this is known as
{\em smoothing}) after forward filtering, then the resulting
mean is equivalent to 4DVAR \cite{fisher2005equivalence}. 
The second observation is that the Bayesian
posterior distribution at the end of the time window, which
is a non-Gaussian version of the Kalman smoothing 
distribution just described, is
equal to the exact filtering distribution at that time, 
provided the filter is initialized with the same distribution 
as that chosen at the start of the time window for the Bayesian
posterior model 
\cite{stuart2010inverse}. The third observation 
is that the 4DVAR variational method corresponds to maximizing
the Bayesian posterior distribution and is known in this
context as a MAP estimator \cite{cox1964estimation, kaipio2005statistical}.
More generally, connections between filtering and smoothing
have been understood for some time \cite{BF63}.

For the filtering and variational algorithms implemented
in practice, these connections may be lost, or weakened,
because of the approximations made to create tractable
algorithms. Hence we attempt to evaluate these algorithms
by their ability to reproduce moments of the Bayesian
posterior distribution since this provides an unequivocal
notion of a perfect solution, given a complete model description,
including sources of error; we hence refer to it as
the gold standard.  We emphasize that we do not claim to 
present optimal implementations of any method except the 
gold standard MCMC.  Nonetheless, the phenomena we 
observe and the conclusions we arrive at will not change 
qualitatively if the algorithms are optimized. They
reflect inherent properties of the approximations
used to create online algorithms useable in practical
online scenarios.

The ability of
filters to track the signal in chaotic systems
has been the object of study in data assimilation
communities for some time and we point to the
paper \cite{miller1994advanced} as an early example
of this work, confined to low dimensional
systems, and to the more recent \cite{carrassi2008data}
for study of both low and high dimensional problems,
and for further discussion of the relevant literature. 
As mentioned above,
we develop our evaluation in the context of the 2D Navier Stokes equations 
in a periodic box. We work in parameter regimes
in which at most $O(10^3)$ Fourier modes
are active.  This model has several attractive features. 
For instance, 
it has a unique global attractor with a tunable parameter, the viscosity
(or, equivalently the Reynolds number), which tunes between 
a one-dimensional stable fixed point and very high-dimensional 
strongly chaotic attractor \cite{temam2001navier}.  
As the dimension of the attractor increases,
many scales are present, as one would expect in a model of the 
atmosphere.  
By working with dimensions of size $O(10^3)$
we have a model of significantly higher dimension 
than the typical toy models that 
one encounters in the literature
\cite{lorenz1996predictability, lorenz1963deterministic}.  
Therefore, while the 2D Navier-Stokes equations do not
model atmospheric dynamics, 
we expect the model to exhibit similar 
predictability issues as arise atmospheric models, and this fact, 
together their high dimensionaliy, makes them a useful model with
which to study aspects of atmospheric data assimilation.
However we do recognize the need for follow-up studies
which investigate similar issues for models such as 
Lorenz-96, or quasi-geostrophic models, which can mimic
or model the baroclinic instabilities which drive so much of
atmospheric dynamics.

The primary conclusions of our study are that: (i) 
with appropriate parameter choices,
approximate filters can perform well in 
reproducing the mean of the desired probability distribution;
(ii) however these filters typically perform poorly when attempting to reproduce 
information about covariance as the assumptions underlying
them may not be valid
(iii) this poor performance is compounded by the need to
modify the filters, and their covariance in particular,
in order to induce filter stability and avoid divergence.
Thus, whilst filters can be a useful tool in predicting mean
behaviour, they should be viewed with caution as predictors
of uncertainty.
These conclusions are intrinsic to the algorithms and will
not change if the model is more complex, for example 
resulting from a smaller viscosity in our model.  
We reiterate that these conclusions are based on
our assumption of well-defined initial prior, 
observational error, and hence Bayesian posterior distributions.
Due to the computational cost of computing the latter we look
only at one, initial, interval of observations, but upon our 
assumption, the accuracy over this first interval will limit accuracy
on all subsequent intervals, and they will not become better.
Under the reasonable assumption that the process has finite
correlation time, the initial prior will be forgotten eventually and,
in the present context, this effect would be explored by choosing different
priors coming from approximation of the asymptotic distribution
by some filtering algorithm and/or climatological statistics
and testing the robustness of conclusions, and indeed of the filtering
distribution itself, to changes in prior.  The question of sensitivity
of the results to choice of prior is not addressed here.  We also
restrict our attention here to the perfect model scenario. 

Many comparisons of various versions of these methods have been 
carried out recently.
For example, \cite{meng2010tests, zhang2010inter} compare  
EnKF forecast with 3DVAR and 4DVAR(without updated covariance)
in the Weather Research and Forecasting (WRF) model.  In their 
real-data experiments, they conclude that EnKF and 4DVAR perform better
with respect the Root Mean Square Error (RMSE), while the EnKF
forecast performs better for longer lead times.  This result is
consistent with ours, although it could be explained by 
an improved approximation of the posterior
distribution at each update time.  Our results indicate 4DVAR
could perform better here, as long as the approximate 
filtering distribution of 4DVAR with the propagated Hessian is
used.  Of course this is too expensive in
practice and often a constant 
covariance is used; this
will limit performance in reproducing the statistical
variation of the posterior filtering distribution for 
prior in the next cycle.  This issue is addressed partially in 
\cite{meng2010tests, zhang2012e4dvar}, where EnKF is coupled
to 4DVAR and the covariance comes from the former, while the 
mean is updated by the latter, and the resulting algorithm outperforms 
either of the individual ones in the RMSE sense.  
Two fundamental classes of EnKFs
were compared theoretically  in the large ensemble limit
in \cite{lei2010comparison},
and it was found that the stochastic version (the one we employ here) 
in which observations are perturbed is more robust 
to perturbations in the forecast distribution than the 
deterministic one.
Another interesting comparison was undertaken in 
\cite{hamill2000comparison} in which several 
ensemble filters alternative to EnKF in operational use
are compared with respect to RMSE as well as other diagnostics 
such as rank histograms \cite{anderson1996method}.  We note 
that over long times the RMSE values for the algorithms we consider
are in the same vicinity as the errors between the 
estimators and the truth
that we present at the single filtering time.

The rest of the paper will be organized in the following sections.
First, we introduce the model and inverse problem
in section \ref{model}, then we describe
the various methods used to (approximately) compute 
posterior smoothing and filtering distributions 
in section \ref{methods}. 
Then we describe the results of the numerical simulations in two sections.
The first, section \ref{accuracy}, explores
the accuracy of the filters by 
comparison with the posterior distribution and the truth.
The second, section \ref{stabilitysec}, explains the manifestation of instability
in the filters, describes how they are stabilized, and
studies implications for accuracy.  
We provide a summary and conclusions in 
section \ref{conclusion}.
In the Appendix \ref{numerics} we
describe some details of the numerical methods.

\section{Statement of the Model}
\label{model}

In this section we describe the dynamical model,
and the filtering and smoothing problems which arise
from assimilating data into that model. The discussion
is framed prior to discretization. Details relating
to numerical implementation may be found 
in the Appendix \ref{numerics}.

\subsection{Dynamical Model: Navier-Stokes Equation}
\label{forwardsec}

The dynamical model we will consider is the two-dimensional
incompressible Navier-Stokes equation in a
periodic box with side of length two. 
By projecting into the space of divergence-free velocity fields,
this may be written as a dynamical equation for the 
divergence-free velocity field $u$ with the form
\begin{linenomath*}
\begin{equation}
\frac{du}{dt} + \nu A u + F(u) = f, \quad u(0)=u_0. 
\label{navier_stokes}
\end{equation}
\end{linenomath*}
Here $A$ (known as the Stokes operator)
models the dissipation and acts as a (negative) 
Laplacian on divergence
free fields, $F(u)$ the nonlinearity 
arising from the convective time-derivative
and $f$ the body force, all projected into divergence free
functions. We also work with spatial mean-zero velocity fields
as, in periodic geometries, the mean evolves independently
of the other Fourier modes. See \cite{temam2001navier} for details
concering the formulation of incompressible fluid mechanics
in this notation. 
We let $\cH$ denote the space of square-integrable, periodic and
mean-zero divergence-free functions on the box.
In order that our results are self-contained apart from the
particular choice of model considered, 
we define the map
$\Psi(\cdot; t) : \cH \rightarrow \cH$ so that
the solution of \eqref{navier_stokes} satisfies 
\begin{linenomath*}
\begin{equation}
u(t) = \Psi(u_0;t).
\label{forward}
\end{equation}
\end{linenomath*}

Equation \eqref{navier_stokes} 
has a global attractor and the viscosity parameter $\nu$
tunes between regimes in which the attractor is a single stationary
point, through periodic, quasi-periodic, chaotic, and strongly chaotic 
(the last two being delicate to distinguish between).  These regimes are
characterized by an increasing number of positive Lyapunov exponents,
and hence increasing dimension of the unstable manifold.  In turn,
this results in a system which becomes progressively less predictable. 
This tunability through all predictability regimes, coupled to the 
possibility of high dimensional effective dynamics which can arise
for certain parameter regimes of the PDE,
makes this a useful model with
which to examine some of the issues inherent in atmospheric data assimilation.

\subsection{Inverse Problem}
\label{inverse}

The basic inverse problem which underlies data assimilation is
to estimate the state of the system, given the model dynamics
for the state, together with noisy observations of the state.
In our setting, since the model dynamics are deterministic,
this ammounts to estimating the initial condition from noisy
observations at later times. This is an ill-posed problem
which we regularize by adopting a Bayesian approach to the
problem, imposing a prior Gaussian random field assumption
on the initial condition. Throughout 
it will be useful to define 
$\|\cdot\|_{B}=\|B^{-\frac12}\cdot\|$
for any covariance operator $B$ and we use this  notation 
throughout the paper, 
in particular in the observation space, with
$B=\Gamma$ and in the initial condition space with $B=\cC_0.$

Our prior regularization on the initial state is to assume
\begin{linenomath*}
\begin{equation}
u_0 \sim \mu_0 = \cN(m_0,\cC_0).
\label{prior}
\end{equation}
\end{linenomath*} 
The prior mean $m_0$ is our best guess of the initial state,
before data is aquired (background mean) and the prior
covariance $\cC_0$ (background covariance) regularizes
this by allowing variability with specified magnitude
at different length-scales.
The prior covariance $\cC_0: \cH \rightarrow \cH$
is self-adjoint and positive, and is assumed to have summable eigenvalues,
a condition which is necessary and sufficient for draws from 
this prior to be square integrable.

Now we describe the noisy observations. We observe
only the velocity field, and not the pressure. Let
$\Gamma: \cH \rightarrow \cH$ 
be self-adjoint, positive operators and let
\begin{linenomath*}
\begin{equation}
y_k \sim \cN( u(t_k),\Gamma)
\label{observations}
\end{equation}
\end{linenomath*}
denote noisy observations of the state at time $t_k = k h$
which, for simplicity of exposition only, we have assumed to
be equally spaced.
We assume independence of the observational
noise: $y_k|u_k$ is independent of  $y_j|u_j$ for all $j \ne k$;
and the observational noise is assumed
independent of the initial condition $u_0.$

For simplicity and following convention in the field, we will not
distinguish notationally between the random variable and its
realization, exept in the case of the truth, which will be important
to distinguish by $u^\dagger$ in subsequent sections in which it
will be simulated and known.
The inverse problem consists of estimating the posterior probability
distribution of $u(t)$, given noisy observations $\{ y_k \}_{k=0}^j$, 
with $j \leq J$.  This is referred to as
\begin{itemize}
\item {\it Smoothing} when $t<t_j$;
\item {\it Filtering} when $t=t_j$; 
\item {\it Predicting} when $t>t_j$.
\end{itemize}
Under the assumption that the dynamical model is deterministic, the
smoothing distribution at time $t=0$
can be mapped forward in time to give the exact
filtering distribution, which in turn can be mapped forward in time to
give the exact predicting distribution (and likewise the
filtering distribution mapped backward, if the forward map admits an
inverse, yields the smoothing distribution). 
If the forward map were linear, for instance in the case of the Stokes equation 
($F(u)=0$), then the posterior distribution would be Gaussian as well, and 
could be given in closed form via its
mean and covariance. In the nonlinear 
case, however,
the posterior cannot be summarized through a finite set
of quantities such as mean and covariance and, in theory,
requires infinitely many samples to represent.    
In the language of the previous 
section, as the dimension of the attractor increases
with Reynolds number, the 
nonlinearity begins to dominate the equation, 
the dynamics become less predictable, and the inverse problem 
becomes more difficult. In particular, Gaussian
approximations can become increasingly misleading.
We will see that sufficient nonlinearity for these misleading
effects can arise more than one way, 
via the dynamical model or the observational frequency.

\subsubsection{Smoothing}
\label{smoothingsec}

We start by describing the Bayesian posterior
distribution, and link this to variational methods.
Let $u_k=u(kh)$, $\Psi(u)=\Psi(u;h)$, and $\Psi^k(\cdot) = \Psi(\cdot;kh)$.  
Furthermore, define the conditional measures for $j_1,j_2 \leq J$ 
\begin{linenomath*}
\begin{equation}
\mu_{j_1|j_2}(u_{j_1}) = \bbP(u_{j_1}| \{y_k\}_{k=0}^{j_2}).
\nonumber
\end{equation}
\end{linenomath*}
(For notational convenience we do not distinguish between
a probability distribution and its density, using $\mu$
and $\bbP$ interchangably for both).
The posterior distributions are completely characterized by the 
dynamical model in
Eq. \eqref{forward} and by the random inputs given in Eq. \eqref{observations} and
Eq. \eqref{prior}.

We focus on the posterior distribution
$\mu_{0|J}$ since this probability
distribution, once known, determines $\mu_{j|J}$ for all
$J \geq j \ge 0$ simply by using \eqref{forward} to map
the probability distribution at time $t=0$ into that
arising at any later time $t>0$.
Bayes' rule gives a characterization of $\mu_{0|J}$ 
via the ratio of its density with respect to that
of the prior \footnote{
Note that our observations include data at time $t=0$. Because
the prior is Gaussian and the observational noise is Gaussian
we could alternatively redefine the prior 
to incorporate this data point, 
which can be done in closed form, and redefine the prior; the
observations would then start at time $t=h$.}:
$$\frac{\bbP(u_0|\{y_k\}_{k=0}^J)}
{\bbP(u_0)}
=\frac{\bbP(\{y_k\}_{k=0}^J| u_0)}{\bbP(\{y_k\}_{k=0}^J)} 
$$
so that
\begin{linenomath*}
\begin{equation}
\frac{\mu_{0|J}(u)}{\mu_0(u)}
\propto \exp\{-\Phi(u)\},
\nonumber
\end{equation}
\end{linenomath*}
where
\begin{linenomath*}
\begin{equation}
\Phi(u) = \frac{1}{2} \left ( \sum_{k=0}^J 
||y_{k}-\Psi^k(u)||^2_\Gamma 
\right ). 
\nonumber
\end{equation}
\end{linenomath*} The constant of proportionality is independent of $u$ and
irrelevant for the algorithms that we use below to probe 
the probability distribution $\mu_{0|J}.$
Note that here, and in what follows, $u$
denotes the random variable $u_0$.

Using the fact that the prior $\mu_0$ is Gaussian
it follows that the {\it maximum a posteriori (MAP) estimator} 
of $\mu_{0|J}$ is the minimizer of the functional
\begin{linenomath*}
\begin{equation}
I(u) = \Phi(u) + \frac{1}{2} 
||u-m_0||^2_{\cC_0}. 
\label{energy}
\end{equation}
\end{linenomath*}
We let ${\tilde{m}}_0=\argmin_u I(u),$
that is ${\tilde{m}}_0$ returns the value of $u$ at which $I(u)$
achieves its minimum.
This so-called MAP estimator is, of course, simply the
solution of the 4DVAR strong constraint variational method.
The mathematical formulation of various inverse problems
for the Navier-Stokes equations, justifying the
formal manipulations in this subsection, may be found in
\cite{cotter2009bayesian}.

\subsubsection{Filtering}
\label{filteringsec}

The posterior filtering distribution at time $j$ given all observations up to time
$j$ can also be given in closed form by an application of Bayes'
rule. The prior is taken as the
predicting distribution:
\begin{eqnarray}
\label{predict}
\mu_{j|j-1}(u_j) &=& \int_{\cH} \bbP(u_j|u_{j-1}) \mu_{j-1|j-1}(d u_{j-1})\\
&=& \int_{\cH} \delta(u_j-\Psi(u_{j-1})) \mu_{j-1|j-1}(d u_{j-1}).
\notag
\end{eqnarray}
The $\delta$ function appears because 
the dynamical model is deterministic. 
As we did for smoothing, we can apply Bayes rule to obtain the 
ratio of the density of $\mu_{j|j}$ with respect to $\mu_{j|j-1}$
to obtain
\begin{linenomath*}
\begin{equation}
\frac{\mu_{j|j}(u)}{\mu_{j|j-1}(u)} 
\propto \exp\{-\Phi_j(u)\},
\label{filtering_density}
\end{equation}
\end{linenomath*}
where
\begin{linenomath*}
\begin{equation}
\Phi_j(u) = \frac{1}{2} ||y_{j}-u||^2_\Gamma. 
\label{loglikelihoodf}
\end{equation}
\end{linenomath*}

Together \eqref{predict} and \eqref{filtering_density}
provide an iteration which, at the final
observation time, yields the measure
$\mu_{J|J}.$ As mentioned in the introduction, this
distribution can be obtained by evolving the posterior
smoothing distribution $\mu_{0|J}$ forward in time under the
dynamics given by \eqref{forward}.

\section{Overview of Methods}
\label{methods}

In this section, we provide details
of the various computational methods we use
to obtain
information about the probability distribution on the
state of the system, given observations, in both the smoothing
and filtering contexts.  
To approximate the gold standard, the Bayesian posterior
distribution, we use state-of-the-art Markov chain Monte
Carlo (MCMC) sampling for the inverse problem, to
obtain a large number of samples from the 
posterior distribution, sufficient to represent its mode and
the posterior spread around it. 
We also decribe optimization techniques to compute
the MAP estimator of the posterior density, namely 4DVAR. 
Both the Bayesian posterior sampling and 4DVAR
are based on obtaining information from the smoothing
distribtion from subsection \ref{smoothingsec}. 
Then we describe a variety of filters,
all building on the description of sequential
filtering distributions introduced in 
subsection \ref{filteringsec}, using Gaussian approximations
of one form or another.  
These filters are 3DVAR, the Fourier Diagonal Filter, 
the Extended Kalman filter, and the Ensemble Kalman filter.  
We will refer to these filtering algorithms collectively
as {\em approximate Gaussian filters} to highlight the fact that
they are all derived by imposing a Gaussian approximation in
the prediction step. 

\subsection{MCMC Sampling of the Posterior}
\label{mcmc}

We work in the setting of the Metropolis-Hastings
variant of MCMC methods, employing recently
developed methods which scale well with respect to
system dimension; see \cite{CDS11} for further
details and references.
The resulting random walk method that we use to sample 
from $\mu_{0|J}$ is given as follows\footnote{w.p. denotes ``with probability''}:
\begin{itemize}
\item Draw $u^{(0)} \sim \cN(m_0,\cC_0)$ and set $n=1.$
\item Define $m^{*}=\sqrt{1-\beta^2} u^{(n-1)}
+(1-\sqrt{1-\beta^2})m_0.$
\item Draw 
$$u^{*} \sim \cN(m^*,\beta^2 \cC_0),$$ 
\item Let $\alpha^{(n-1)}=\min\Bigl\{1,\exp\bigl(\Phi(u^{(n-1)})-\Phi(u^{*})
\bigr)\Bigr\}$ and set
\begin{eqnarray*}
u^{(n)} = \left\{
\begin{array}{cc}
 u^{*} &  {\rm w.p.}
\,\alpha^{(n-1)}\\
  u^{(n-1)} & else. 
\end{array}
\right \}
\end{eqnarray*}
\item $n \mapsto n+1$ and repeat.
\end{itemize}

After a burn-in period of $M$ steps, $\{u^{(n)}\}_{n=M}^N \sim \mu_{0|J}.$ 
This sample is then pushed forward to yield a sample of 
time-dependent solutions, 
$\{u^{(n)}(t)\}$, where $u^{(n)}(t)=\Psi(u^{(n)};t)$, 
or in particular in what follows, 
a sample of the filtering distribution $\{\Psi^J u^{(n)}\}$.

\subsection{Variational Methods: 4DVAR}
\label{var4}

As described in section \ref{model}, the
minimizer of $I$ defined in Eq. \eqref{energy} defines
the 4DVAR approximation, the basic variational
method.  
A variety of optimization routines can be used to solve
this problem. We have found Newton's method to
be effective, with an initial starting point computed
by homotopy methods starting from an easily computable
problem.

We now outline how the 4DVAR solution may be used to
generate an approximation to the distribution of
interest. The 4DVAR solution (MAP estimator) coincides with the 
mean for unimodal symmetric distributions.  
If the variance under $\mu_{0|J}$ is small
then it is natural to seek a Gaussian approximation. This
has the form $\cN(\tilde{m}_0,\tilde{\cC}_0)$ where
$$\tilde{\cC}_0^{-1} = D^2I(\tilde{m}_0) = D^2\Phi(\tilde{m}_0) + {\cC}_0^{-1}.$$
Here $D^2$ denotes the second derivative operator.
This Gaussian on the initial condition $u_0$
can be mapped forward under the dynamics, using
linearization for the covariance, since it is assumed small,
to obtain $u(t) \approx \cN\bigl({\tilde m}(t),{\tilde \cC}(t)\bigr)$ where
${\tilde m}(t)=\Psi(\tilde{m}_0;t)$  and
$${\tilde \cC}(t) = D\Psi(\tilde{m}_0;t) \tilde{\cC}_0 D \Psi(\tilde{m}_0;t)^*.$$
Here $D$ denotes the derivative operator, and $^*$ the adjoint.

\subsection{Approximate Gaussian Filters}
\label{varfilters}

Recall the key update formulae \eqref{predict},
\eqref{filtering_density}.
Note that the integrals are over the function space $\cH$, a fact
which points to the extreme computational complexity of characterizing
probability distributions for problems arising from PDEs or their
high dimensional approximation. We will describe various 
approximations, 
which are all Gaussian in nature, and make the update
formulae tractable.
We describe some generalities relating to this issue,
before describing various method dependent specifics
in following subsections.

If $\Psi$ is nonlinear then 
$\mu_{j-1|j-1}$ Gaussian does not imply $\mu_{j|j-1}$ is
Gaussian; this follows from \eqref{predict}.
Thus prediction cannot be performed simply by mapping
mean and covariance. 
However, the update equation \eqref{filtering_density} has the
property that, if $\mu_{j|j-1}$ is Gaussian then so is 
$\mu_{j|j}.$ 
If we assume that $\mu_{j|j-1}=\cN(m_j , \cC_j)$, 
then \eqref{filtering_density} shows that
$\mu_{j|j}$ is Gaussian $\cN({\hat m}_j,{\hat \cC}_j)$
where ${\hat m}_j$ is the MAP estimator given by 
\begin{linenomath*}
\begin{equation}
{\hat m}_j=\argmin_u I_j(u),
\label{energyf}
\end{equation}
\end{linenomath*}
(so that ${\hat m}_j$ minimizes $I_j(u)$)
and
$$I_j(u) = \Phi_j(u) + \frac{1}{2} ||u-m_j||^2_{\cC_j}.$$ 
Note that, using \eqref{loglikelihoodf},
we see that $I_j$ 
is a quadratic form whose minimizer is given in 
closed form as the solution of a linear equation with the form
\begin{linenomath*}
\begin{equation}
\label{eq:mean}
{\hat m}_j={\hat \cC}_j\Bigl(\cC_j^{-1}m_j
+\Gamma^{-1}y_j\Bigr)
\end{equation}
\end{linenomath*}
where
\begin{linenomath*}
\begin{equation}
\label{eq:cov}
{\hat \cC}_j^{-1}=\cC_j^{-1}+\Gamma^{-1}.
\end{equation}
\end{linenomath*}
If the output of the prediction step given by \eqref{predict} 
is approximated
by a Gaussian 
then this provides the basis for a
sequential Gaussian approximation method.
To be precise, if we have that 
$$\mu_{j-1|j-1}=\cN({\hat m}_{j-1},{\hat \cC}_{j-1})$$ 
and we have formulae, based on an approximation
of \eqref{predict}, which enable us to compute the map
\begin{linenomath*}
\begin{equation}
\label{forward2}
({\hat m}_{j-1},{\hat \cC}_{j-1}) \mapsto (m_j,\cC_j)
\end{equation}
\end{linenomath*} 
then together \eqref{eq:mean}, \eqref{eq:cov}, \eqref{forward2}
provide an iteration for Gaussian approximations of
the filtering distribution $\mu_{j|j}$ of the form
\begin{linenomath*}
\begin{equation}
\nonumber
({\hat m}_{j-1},{\hat \cC}_{j-1}) \mapsto 
({\hat m}_{j},{\hat \cC}_{j}). 
\end{equation}
\end{linenomath*} 
In the next few subsections
we explain a variety of such approximations,
and the resulting filters.

\subsubsection{Constant Gaussian filter (3DVAR)}
\label{var3}

The constant Gaussian filter, referred to as 3DVAR, consists 
of making the choices
$m_{j}= \Psi(\hat{m}_{j-1})$ and $\cC_j \equiv \cC$
in \eqref{forward2}. It is natural, theoretically,
to choose $\cC=\cC_0$
the prior covariance on the initial condition. However, as we will
see, other issues may intervene and suggest or necessitate other
choices.

\subsubsection{Fourier Diagonal Filter (FDF)}
\label{fdf}

A first step beyond 3DVAR, which employs constant covariances
when updating to incorporate new data, is 
to use some approximate dynamics in order 
to make the update \eqref{forward2}. 
In \cite{harlim2008filtering,majda2010mathematical} it is demonstrated
that, in regimes exhibiting chaotic dynamics,
linear stochastic models can be quite effective
for this purpose: this is the idea of  
the Fourier Diagonal Filter. In this subsection we
describe  how this idea may be used,
in both the steady and trubulent regimes of the
Navier-Stokes system under consideration. 
For our purposes, and as observed in \cite{harlim2008filtering},
this approach provides a rational way of deriving the
covariances in 3DVAR, based on climatological statistics.

The basic idea is, for the purposes of filtering,
to replace the nonlinear map
$u_{j+1}=\Psi(u_j)$ by the linear (stochastic 
when $\cQ \ne 0$) map 
\begin{linenomath*}
\begin{equation}
u_{j+1} = Lu_j + \sqrt{\cQ}\xi_j.
\label{OUforward}
\end{equation}
\end{linenomath*}
Here it is assumed that
$L$ is negative definite and diagonal in the
Fourier basis, $\cQ$ has summable eigenvalues and is diagonal
in the Fourier basis and $\xi_j$ is a random noise
chosen from the distribution $\cN(0,I)$.  
More sophisticated linear stochastic models could
(and should) be used, but we employ this simplest of
models to convey our ideas.

If $L=\exp(-M h)$ and $\cQ=[I-\exp(-2M h)] \Xi$, 
then \eqref{OUforward} corresponds to 
the discrete time $h$ solution of the 
Ornstein-Uhlenbeck (OU) process
\begin{linenomath*}
\begin{equation}
du + M u dt = \sqrt{2 M\Xi} dW,
\nonumber
\end{equation}
\end{linenomath*}
where $dW$ is the infinitesimal Brownian motion increment
with identity covariance.
The stationary solution is $\cN(0,\Xi)$ and letting
$M_{k,k}=\alpha_k$, 
the correlation time for mode $k$ can be computed as 
$1/\alpha_k$.  
We employ three models of the form \eqref{OUforward} in this
paper, labelled a), b) and c), and detailed below.
Before turning to them, we describe how this linear
model is incorporated into the filter.

In the case of linear dynamics such as these, the
map \eqref{forward2} is given in closed form
\begin{equation*}
m_j=L\hat{m}_{j-1}, \quad \cC_j=L \hat{\cC}_{j-1}L^* +\cQ. 
\nonumber
\end{equation*}
This can be improved, however, in the spirit of 3DVAR, by updating 
only the covariance in this way and mapping the mean under
the nonlinear map, to obtain the following instance of
\eqref{forward2}:
\begin{linenomath*}
\begin{equation}
m_j=\Psi(\hat{m}_{j-1}),\quad \cC_j=L \hat{\cC}_{j-1}L^* +\cQ. 
\nonumber
\end{equation}
\end{linenomath*}
We implement the method in this form. We note that,
because $L$ is negative-definite, the
covariance $\cC_j$ 
converges to some $\cC_{\infty}$ which can be computed explicitly,
and, asymptotically, the algorithm behaves like 3DVAR 
with a systematic choice of covariance.
We now turn to the choices of $L$ and $\cQ$.

{\bf Model (a)} is used in the stationary regime.
It is found by setting $L=\exp(-\nu Ah)$ and taking $\cQ=\epsilon I$
where $\epsilon=10^{-12}$. Although this
does not correspond to an accurate linearization of
the model in low wave numbers, it is reasonable for high wave
numbers.

{\bf Model (b)} is used in the strongly chaotic regime, and
is based on the original idea in 
\cite{harlim2008filtering,majda2010mathematical}. 
The two quantities $\Xi_{k,k}$ and $\alpha_k$ are matched to
the statistics of the dynamical model, as 
follows.  Let $u(t)$ denote the solution to the 
Navier-Stokes equation (\ref{navier_stokes}) which,
abusing notation, we assume to be represented in the
Fourier domain, with entries $u_k(t).$  Then  
$\bar{u}$ and $\Xi$ are given by the formulae
\begin{align*}
\bar{u} &=\lim_{T\rightarrow \infty}\frac{1}{T}\int_0^T u(t) dt,\\
\Xi&=  \lim_{T\rightarrow \infty}\frac{1}{T}\int_0^T (u(t)-\bar{u}) 
\otimes (u(t)-\bar{u})^* dt. 
\end{align*}
In practice these integrals are 
approximated by finite discrete sums. Furthermore, we set
the off-diagonal entries of $\Xi$ to zero to obtain
a diagonal model.  We set $\sigma_k^2=\Xi_{k,k}.$ Then
the $\alpha_k$ are computed using the formulae
\begin{eqnarray*}
M(t,\tau)&=(u(t-\tau)-\bar{u}) \otimes (u(t)-\bar{u})^*\\
{\rm Corr}_k(\tau)&=
\lim_{T\rightarrow \infty}\frac{1}{\sigma_k^2}\int_0^T M_{k,k}(t,\tau) dt\\ 
\alpha_k &= \Bigl(\int_0^{\infty} {\rm Re(Corr}_k(\tau)) d\tau\Bigr)^{-1}.
\end{eqnarray*}
Again, finite discrete sums are used to approximate the
integrals.



\subsubsection{Low Rank Extended Kalman Filter (LRExKF)} 
\label{lrexkf}

The idea of the extended Kalman filter is to assume
that the desired distributions are approximately Gaussian
with small covariance. Then linearization may be used to
show that a natural approximation of \eqref{forward2} is 
the map \footnote{As an aside, we note a more
  sophisticated 
improved version we have not seen yet in the literature would 
include the higher-order drift term involving the Hessian.
Although adding significant expense
there could be scenarios in which this is worthwhile 
to attempt this.}
\begin{linenomath*}
\begin{equation}
m_j=\Psi(\hat{m}_{j-1}),\quad 
\cC_j = D\Psi(\hat{m}_{j-1})
\hat{\cC}_{j-1}D\Psi(\hat{m}_{j-1})^*.
\label{covariance}
\end{equation}
\end{linenomath*}
Updating the covariance this way
requires one forward tangent linear 
solve and one adjoint solve for each dimension of the system, and
is therefore prohibitively expensive for high
dimensional problems.  
To overcome this
we use a low rank approximation to the covariance update.

We write this explicitly as follows.
Compute the dominant $m$ eigenpairs of $\cC_j$ as
defined in Eq. \eqref{covariance}; these satisfy 
$$
D\Psi(\hat{m}_{j-1})\hat{\cC}_{j-1}D\Psi(\hat{m}_{j-1})^* V =  V \Lambda
$$
Define the rank $m$ matrix $\cM = V \Lambda V^*$ and 
note that this captures the essence of
the covariance implied by the extended Kalman filter, in the
directions of the $m$ dominant eigenpairs.
When the eigenvalues are well-separated, as they
are here, a small number of eigenvalues capture the majority 
of the action and this is very efficient.
We then implement the filter
\begin{linenomath*}
\begin{equation}
m_j=\Psi(\hat{m}_{j-1}),\quad 
\cC_j = \cM +  \epsilon I
\label{covariance2}
\end{equation}
\end{linenomath*}
where $\epsilon=10^{-12}$ as above.  
The perturbation term prevents degeneracy.

The notion of keeping track of the unstable directions of the 
dynamical model is not new, although our particular implementation
differs in some details. For discussions and examples of this
idea see \cite{toth1997ensemble}, \cite{palmer1998singular}, 
\cite{kalnay2003atmospheric}, \cite{leutbecher2003adaptive}, 
\cite{auvinen2009large}, and \cite{hamill2000comparison}.

\subsubsection{Ensemble Kalman Filter (EnKF)}
\label{enkf}

The Ensemble Kalman Filter, introduced in
\cite{evensen1994assimilation} and overviewed in
\cite{evensen2003ensemble, evensen2009data}, is slightly outside the framework of the
previous three filters
and there are many versions (see \cite{lei2010comparison}
  for a comparison between two major categories.) 
This is because the basic object
which is updated is an ensemble of particles, not a mean
and covariance. This ensemble is used to compute an empirical
mean and convariance. We describe how the basic building
blocks of approximate Gaussian filters, namely \eqref{eq:mean}, 
\eqref{eq:cov} and \eqref{forward2}, 
are modified to use ensemble statistics.

We start with \eqref{forward2}.
Assuming one has an ensemble 
$\{\hat{m}_{j-1}^{(n)}\} \sim \cN(\hat{m}_{j-1},\hat{\cC}_{j-1})$,
\eqref{forward2} is replaced by the approximations 
\begin{align*}
m_j^{(n)}&=\Psi(\hat{m}^{(n)}_{j-1}) \notag\\
m_j&= \frac{1}{N} \sum_{n=1}^{N} m_j^{(n)} 
\label{enkf:mean}
\end{align*}
and
\begin{linenomath*}
\begin{equation}
\cC_j= \frac{1}{N} \sum_{n=1}^{N} (m_j^{(n)}-m_j) (m_j^{(n)}-m_j)^*.
\label{enkf:cov}
\end{equation}
\end{linenomath*}
The equation \eqref{eq:mean} is approximated via an ensemble
of equations found by replacing $m_j$ by $m_j^{(n)}$
and replacing $y_j$ by independent draws 
$\{y_j^{(n)}\}$ from $\cN(y_j,\Gamma).$
This leads to updates of the
ensemble members $m_j^{(n)} \mapsto {\hat m}_j^{(n)}$
whose sample mean yields $\hat{m}_j$. 
For infinite particles, the sample covariance 
yields $\hat{\cC}_j$.  In the comparisons we consider 
the covariance to be the analytical one 
$\hat{\cC}_j = (I - \cC_{j-1} (\cC_{j-1}+\Gamma)^{-1}) \cC_{j-1}$
as in \eqref{eq:cov}, rather than the ensemble
sample covariance, which yields the one implicitly in the next update
\eqref{forward2}.  The discrepancy between these can
be large for small samples and in different situations 
it may have either a positive or 
negative effect on the filter divergence discussed in Section 
\ref{stabilitysec}.
Solutions of the ensemble of equations
of form \eqref{eq:mean} are implemented in the standard
Kalman filter fashion; this does not involve computing
the inverse covariances which appear in \eqref{eq:cov}.
There are many variants on the EnKF and we have
simply chosen one representative version. 
See, for example, \cite{tippett2003ensemble} and 
\cite{evensen2009data}.

\section{Filter Accuracy}
\label{accuracy}

In this section we describe various aspects of the accuracy of 
both variational methods (4DVAR) and approximate
Gaussian filters, evaluating them with
respect to their effectiveness in reproducing the following 
two quantities: (i) the
posterior distribution on state given observations; (ii) 
the truth $u^{\dagger}$ which gives rise to the
observations.  
The first of these is found by means of accurate
MCMC simulations, and is then characterized by three
quantities: its mean, variance, and MAP estimator.  
It is our contention that, where quantification of uncertainty
is important, the comparison of algorithms by their ability
to predict (i) is central; however many algorithms are
benchmarked in the literature by their ability to
predict the truth (ii) and so we also include this information.
A comparison of the algorithms with (iii) the observational 
data is also included as a useful
check on the performance of the algorithms. Note that
studying the error in (i) requires comparison of
probability distributions; we do this primarily through
comparison of mean and covariance information. In all our
simulations the posterior distribution, and the distributions
implied by the variational and filtering algorithms,
are approximately Gaussian; for this reason studying
the mean and covariance is sufficient.
We note that we have not included model error in our study:
uncertainty in the dynamical model comes only through the
initial condition; thus attempting to match the ``truth'' 
is not unnatural in our setting. Matching the posterior
distribution is, however, arguably more natural and is a concept
which generalizes in a straightforward fashion
to the inclusion of model error.  In this section all methods are
presented in their ``raw'' form, unmodified and not optimized.  
Modifications that are often used
in practice are discussed in the next section.

\subsection{Nature of Approximations}
\label{nature}

In this preliminary discussion we make {\em three
observations} which help to guide and understand
subsequent numerical experiments. 
For the purposes of this discussion we assume that
the MCMC method, our gold standard, provides {\em exact} 
samples from the desired posterior distribution.
There are then two key approximations underlying the methods
which we benchmark against MCMC in this section. 
The first is the
Gaussian approximation, which is made in 3DVAR/FDF, 4DVAR (when
propagating from $t=0$ to $t=T$), LRExKF and EnKF; the
second additional approximation is sampling, 
which is made only in EnKF. 
The 3DVAR and FDF methods make a universal, steady approximation
to the covariance whilst 4DVAR, LRExKF and EnKF all
propagate the approximate covariance using the dynamical
model. Our first observation is thus that 
{\em we expect 3DVAR and FDF to underperform
the other methods with regard to covariance information.}
The second observation arises from the following:
the predicting (and hence smoothing and filtering) distribution will
remain close to Gaussian as long as 
there is a balance between dynamics remaining close to linear 
and the covariance being small enough (i.e. there is an appropriate
level of either of these factors which can counteract any instance of 
the other one).  
In this case the evolution of the distribution is well approximated to leading
order by the non-autonomous linear system update of ExKF, and
similarly for the 4DVAR update from $t=0$ to $t=T$.  
Our second observation is hence that
{\em the bias in the Gaussian approximation will
become significant if the dynamics is sufficiently non-linear
or if the covariance becomes large enough}. 
These two factors which destroy the Gaussian approximation
will be more pronounced as the Reynolds number increases,
leading to more, and larger, growing (local) Lyapunov exponents,
and as the time interval between observations increases,
allowing further growth or, for 4DVAR, as the total time-interval
grows. 
The third and final observation concerns EnKF methods.
In addition to making the Gaussian approximation,
these rely on sampling to capture the resulting
Gaussian. Hence the error 
in the EnKF methods will
become significant if the number of samples is too small,
even when the Gaussian approximation is valid.
Furthermore, since the number of samples required tends
to grow with both dimension and with the inverse of the
size of the quantity being measured, we expect that
EnKF will encounter difficulties in this high dimensional
system which
will be exacerbated when the covariance is small.

We will show in the following that 
in the stationary case, and for high frequency observations 
in the strongly chaotic case, the ExKF does perform well because of an 
appropriate balance of the level of nonlinearity of the dynamics 
on the scale of the time between observations and the magnitude 
of the covariance.  Nonetheless, a reasonable sized ensemble in
the EnKF is not sufficiently large for the error from that algorithm 
to be dominated by the ExKF error, and it is instead determined by 
the error in the sample statistics with which EnKF approximates 
the mean and covariance; this latter effect was demonstrated
on a simpler model problem in \cite{apte2008data}.
When the observations are sufficiently sparse in time in the 
strongly chaotic case the Gaussian approximation is no longer valid 
and even the ExKF fails to recover accurate mean and covariance.

\subsection{Illustration via Two Regimes}
\label{illustration}

This section is divided into two subsections, 
each devoted to a dynamical regime:
stationary, 
and strongly chaotic.  The true initial condition $u^{\dagger}$ in the case of strongly chaotic
dynamics is taken as an arbitrary point on the attractor obtained by
simulating an arbitrary initial condition until statistical
equilibrium.  The initial condition for the case of stationary
dynamics is taken as a draw from the Gaussian prior, 
since the statistical equilibrium is the trivial one. 
Note that in the stationary dynamical regime the 
equation is dominated by the linear term and hence this regime acts
as a benchmark for the approximate Kalman filters, since they 
are exact in the linear case.
Each of these sections 
in turn explores the particular characteristics of the filter accuracy
inherent to that regime as a function of time between observations, $h$.
The final time, $T$, will mostly be fixed, so that decreasing
$h$ will increase the density of observations of the
system on a fixed time domain; however, on several occasions
we study the effect of fixing $h$ and changing the final
time $T$. Studies of the effect on the posterior distribution
of increasing the number of observations are
undertaken for some simple inverse problems in fluid
mechanics in \cite{CDS11} and are not undertaken here. 

We now explain the basic format of the tables which follow and
indicate the major features of the filters that they
exhibit.  The first $8$ rows 
each correspond to a {\em method}
of assimilation, while the final two rows correspond to
the truth, at the start and end of the time window
studied, for completeness. Labels for these rows
are given in the far left column.  
The posterior distribution (MCMC) and MAP
estimator (4DVAR) are each obtained via the smoothing distribution,
and hence comparson is made at the intial time, $t=0$,
and at the final time, $t=T$, by mapping forward.  
For all other methods, the
comparison is only with the filtering distribution at the final time, $t=T$.
The columns each indicate the
relative error of the given filter with a particular diagnostic quantity of interest.
The first, third, fourth and fifth columns show 
$e=||M(t)-m(t)||/||M(t)||$, 
where $M$ is, respectively, the mean of the 
posterior distribution
found by MCMC and denoted $\bbE u(t)$,
the truth $u^\dagger(t)$, the observation $y(t)$, 
or the MAP estimator (4DVAR) at time $t$ (either $0$ or $T$) 
and $m(t)$ is the time $t$ mean of the filtering (or smoothing)
distribution obtained from each of the various methods.  
The norm used is the $L^2\bigl([-1,1)\times[-1,1)\bigr)$
norm. The second column shows 
$$e=\frac{\|{\rm var}\bigl(u(t)\bigr)-{\rm var}\bigl(U(t)\bigr)\|}
{\|{\rm var}\bigl(u(t)\bigr)\|}$$
where ${\rm var}$ indicates the variance, $u$ is sampled
from the posterior distribution (via MCMC), 
and $U$ is the Gaussian approximate state obtained from the various methods.
The subscripts in the titles in the top row indicate 
which relative error is given in that column.

The following universal observations can be made independent of 
model parametric regime.

\begin{itemize}

\item The numerical results support the three
observations made in the previous subsection.

\item Most algorithms do a reasonably god job of reproducing
the mean of the posterior distribution.

\item The LRExKF and 4DVAR both do a 
reasonably good job of 
reproducing the variance of the posterior distribution if
the Reynolds number is sufficiently small and/or the 
observation frequency high; otherwise 
there are circumstances
where the approximations underlying the ad hoc filters are
not justified and they then fail to reproduce covariance information
with any accuracy.

\item All other algorithms perform poorly when reproducing the
variance of the posterior distribution.

\item All estimators of the mean
are uniformly closer to the truth than the
observations for all $h$.

\item In almost all cases, the estimators of the mean
are closer to the mean of
the posterior distribution than to the truth.

\item The error of the estimators of the mean with respect 
to the truth tends to increase with increasing $h$.  

\item The error of the mean with respect to the truth decreases 
for increasing number of observations. 

\item LRExKF usually has the smallest
error with respect to the posterior mean and 
sometimes
accurately recovers the variance. 

\item The error in the variance is sometimes overestimated
and sometimes underestimated, and usually this is 
wavenumber-dependent in the sense that the variance of 
certain modes is overestimated and the variance of others 
is under-estimated.  This will be discussed further in the next section.

\item The posterior smoothing distribution becomes noticeably 
non-Gaussian although still unimodal, while the filtering distribution 
remains very close to Gaussian.

\end{itemize}

\subsection{Stationary Regime}
\label{stationaryr}

\begin{figure*}
\includegraphics[width=1\textwidth]{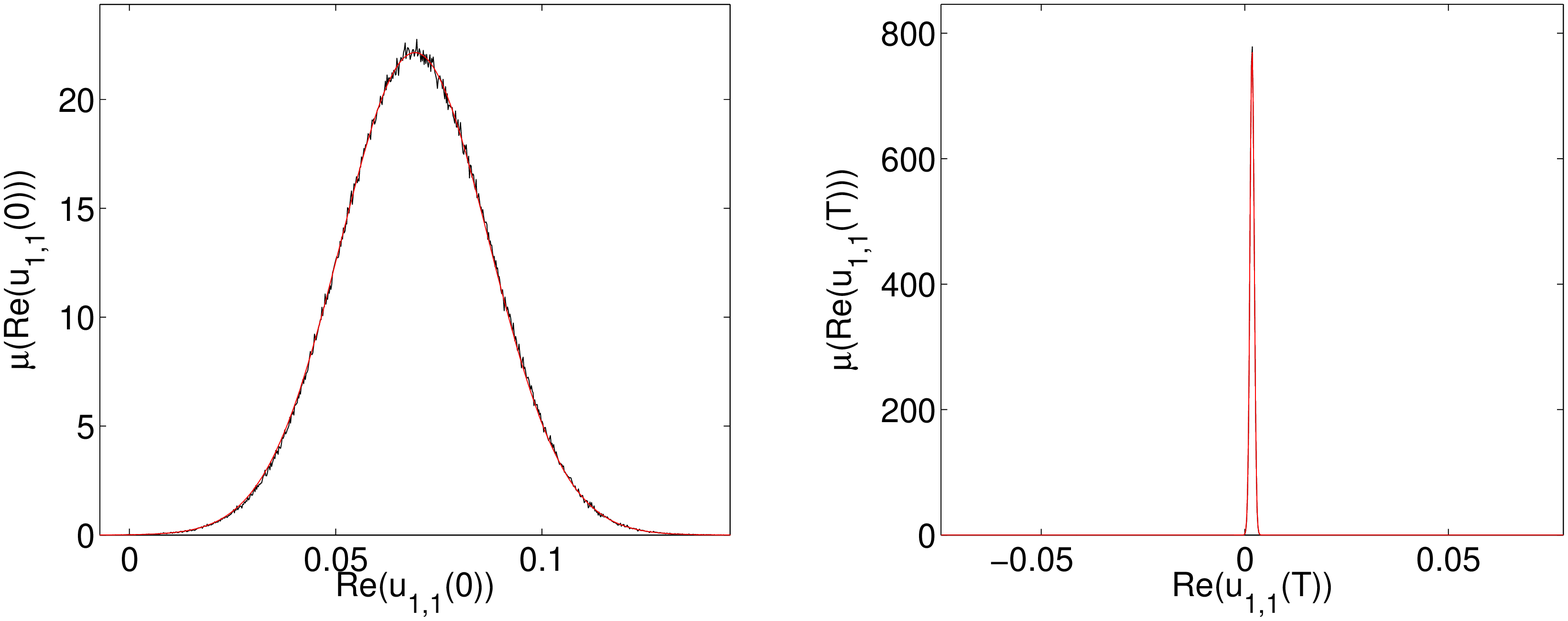}
\caption{The MCMC histogram at $t=0$ (left) and $t=T=10h=2$ (right)
together with the Gaussian approximation obtained from 4DVAR for
low Reynolds number, stationary state regime ($\nu=0.1$).}
\label{stationaryv}
\end{figure*}

In the stationary regime, $\nu=0.1$, the basic time-step used is 
$dt=0.05$, the smallest $h$ considered is $h=0.2$, and we fix 
$T=2$ as the filtering time at which to make comparisons of the
approximate filters with 
the moments of the posterior distribution via samples from 
MCMC, the MAP estimator from 4DVAR, the truth, 
and the observations.  Figure \ref{stationary1} shows the 
vorticity, $w$ (left), and Fourier coefficients, $|u_k|$ (right), of the smoothing
distribution at $t=0$ in the case that $h=0.2$.  
The top panels are the mean of the posterior 
distribution found with MCMC, ($\bbE u)$, and 
the bottom panels are the truth, $u^\dagger(0)$.
The MAP estimator (minimizer
of $I(u)$, 
$\hat{m}_0={\rm argmin}\, I$)
is not shown because it is not discernable from the mean 
in this case.
Notice that the mean (and MAP estimator) on the initial
condition 
resemble the large-scale structure of the truth, but are more 
rough.  This roughness is caused by the
presence of the prior mean $m_0$ drawn 
according to the distribution $\cN(u^\dagger(0),\cC_0)$. 
The solution operator $\Psi$ immediately removes this roughness
as it damps high wavenumbers; this effect 
can be seen in the images of the smoothing distribution 
mapped forward to time $t=T$, i.e. the filtering distribution, 
in Figure \ref{stationary2} (
here only the mean is shown, as neither the truth
nor the MAP estimator are distinguishable from it).
This is
apparent in the data in the tables discussed
below, in which the distance between
the truth, the posterior distribution, and the MAP estimator are all
mutually much closer for the final time than the initial;
this contraction of the errors in time is caused
by the underlying dynamics which involves exponential
attraction to  a unique stationary state. 
This is further exihibited in
Figure \ref{stationaryv} which shows 
the histogram of the smoothing distribution for
the real part of a sample mode, $u_{1,1}$, 
at the initial time (left) and final time
(right).

Table \ref{stationaryh1} 
presents data for increasing $h=0.2, 1, 2$, with $T=2$
fixed.  Notable trends, in addition to those
mentioned at the start of this section, are as follows:
(i) the 4DVAR smoothing distribution has much smaller error with respect
  to the mean at $t=T$
than at $t=0$, with the former increasing and the latter decreasing 
for increasing $h$; 
the error of 4DVAR with respect to the mean and the 
variance at $t=0$ and $t=T$ 
are close to or below the threshold of accuracy of MCMC;
(iii) the error of both the mean and the variance of 3DVAR 
tend to decrease with increasing $h$;

\subsection{ Strongly Chaotic Regime}
\label{turbulentr}

\begin{figure*}
\includegraphics[width=0.5\textwidth]{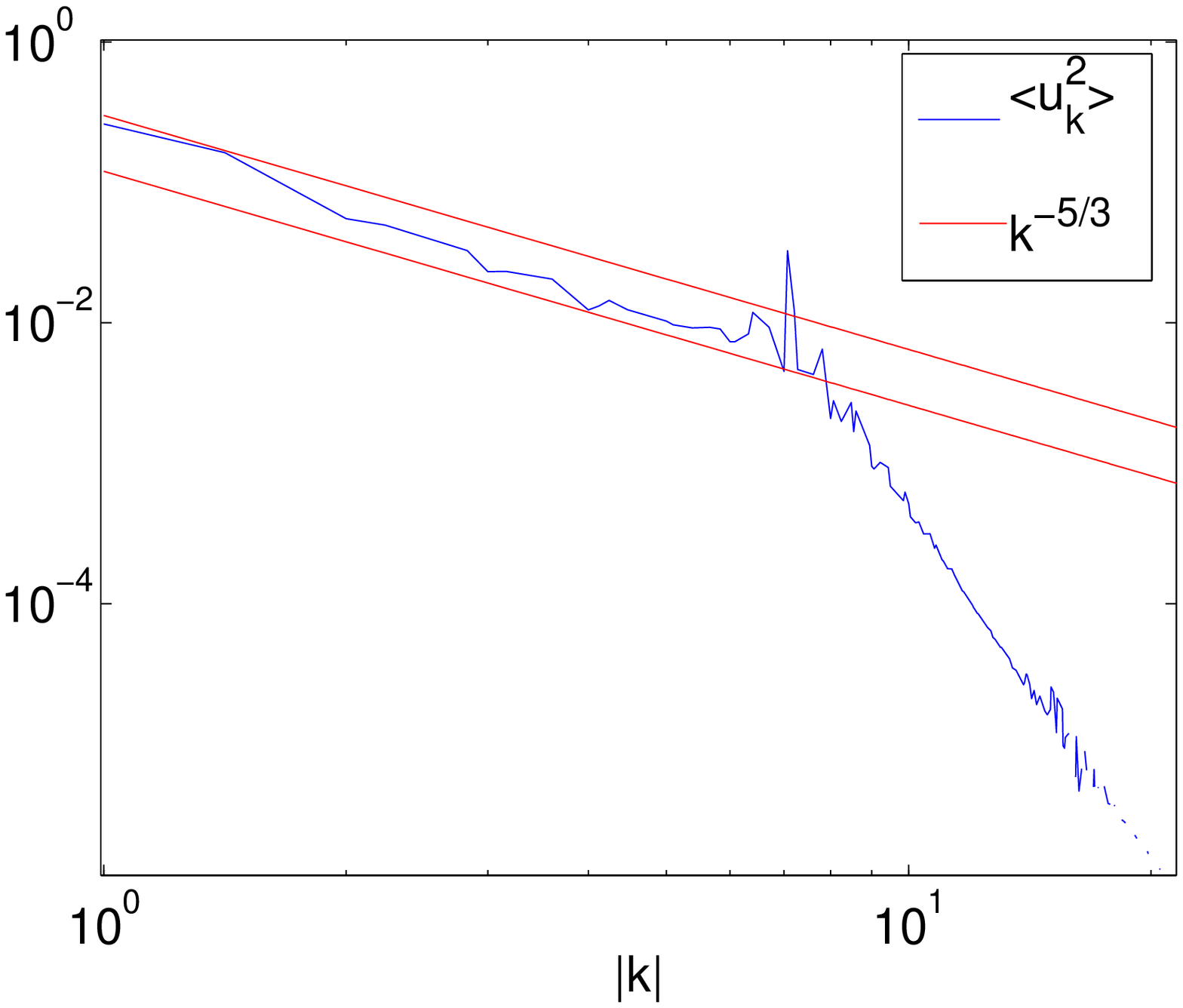}
\includegraphics[width=0.5\textwidth]{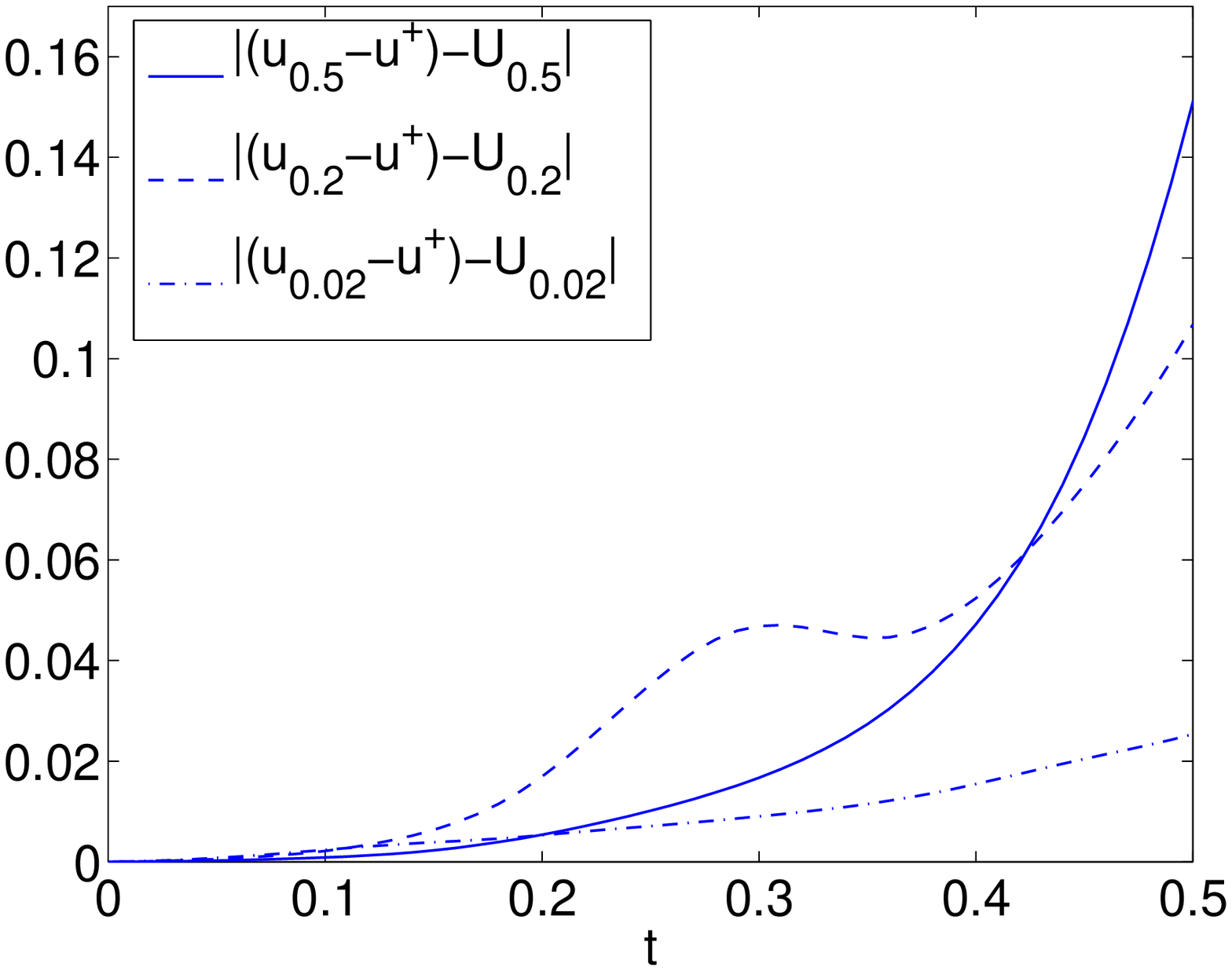}
\caption{The left panel is the 
average velocity spectrum on the attractor for $\nu=0.01$. 
The right panel shows the difference between (a) and (b)
where: (a) is the difference
of the truth $u^{\dagger}(t)$ with a solution $u_\tau(t)$ 
initially perturbed in the direction of the dominant 
local Lyapunov vectors $v_\tau$, on time-interval of length $\tau$, 
with $\tau=0.02, 0.2$, and $0.5$
(thus $u_\tau(0) = u^{\dagger}(0) + \varepsilon v_\tau$);
and (b) is the evolution of that perturbation under the 
linearized model $U_{\tau}(t)=D\Psi(u^{\dagger}(0);t) \varepsilon v_\tau$.  
The magnitude of perturbation $\varepsilon$ is determined by
the projection of the initial posterior covariance in the 
direction $v_\tau$. The difference plotted thus indicates 
differences between linear and nonlinear
evolution with the the direction of the initial perturbations
chosen to maximize growth and with size of the initial
perturbations commensurate with the prevalent uncertainty.
The relative error
$|[u_\tau(\tau) - u^\dagger(\tau)]-U_\tau(\tau)|/|U_\tau(\tau)|$
(in $l^2$)
is $0.01, 0.15,$ and $0.42$, respectively, for the three chosen
values of increasing $\tau$. }
\label{spectrum}
\end{figure*}

In the strongly chaotic regime, $\nu=0.01$, the basic time-step used is 
$dt=0.005$, the smallest $h$ considered is $h=0.02$, and we fix 
$T=0.2$ or $T=1$ 
as the filtering time at which to make 
comparisons of the approximate filters.  
In this regime, the dynamics are significantly 
more nonlinear and less predictable, with a high-dimensional 
attractor spanning many scales.  
Indeed the energy spectrum decays like $E(k)  = \lim_{\delta
  \rightarrow 0} \int_{0}^{2\pi} \int_k^{k+\delta} \bbE |u_l|^2 l dl d\theta \propto k^{-2/3}$
for $|k|<k_f$, with $k_f$ the magnitude of the forcing wavenumber,
and much more rapidly for $|k|>k_f$.  
See the left panel of Figure \ref{spectrum} for the
average spectrum of the solution on the attractor and Fig. \ref{turb1} for an example
snapshot of the solution on the attractor.  
The flow is not in any of
the classical regimes of cascades, but 
there is an upscale transfer of 
energy because of the forcing at intermediate scale. 
The viscosity is not negligible even at the largest scales, thereby allowing
statistical equilibrium; this may be thought of
as being generated by the empirical measure on the global
attractor whose existence is assured for all $\nu>0$.  
We confirmed this with simulations to times of
order $O(10^3 \nu)$ giving $O(10^7)$ samples with which to compute
the converged correlation statistics used in FDF.

\begin{figure*}
\includegraphics[width=1\textwidth]{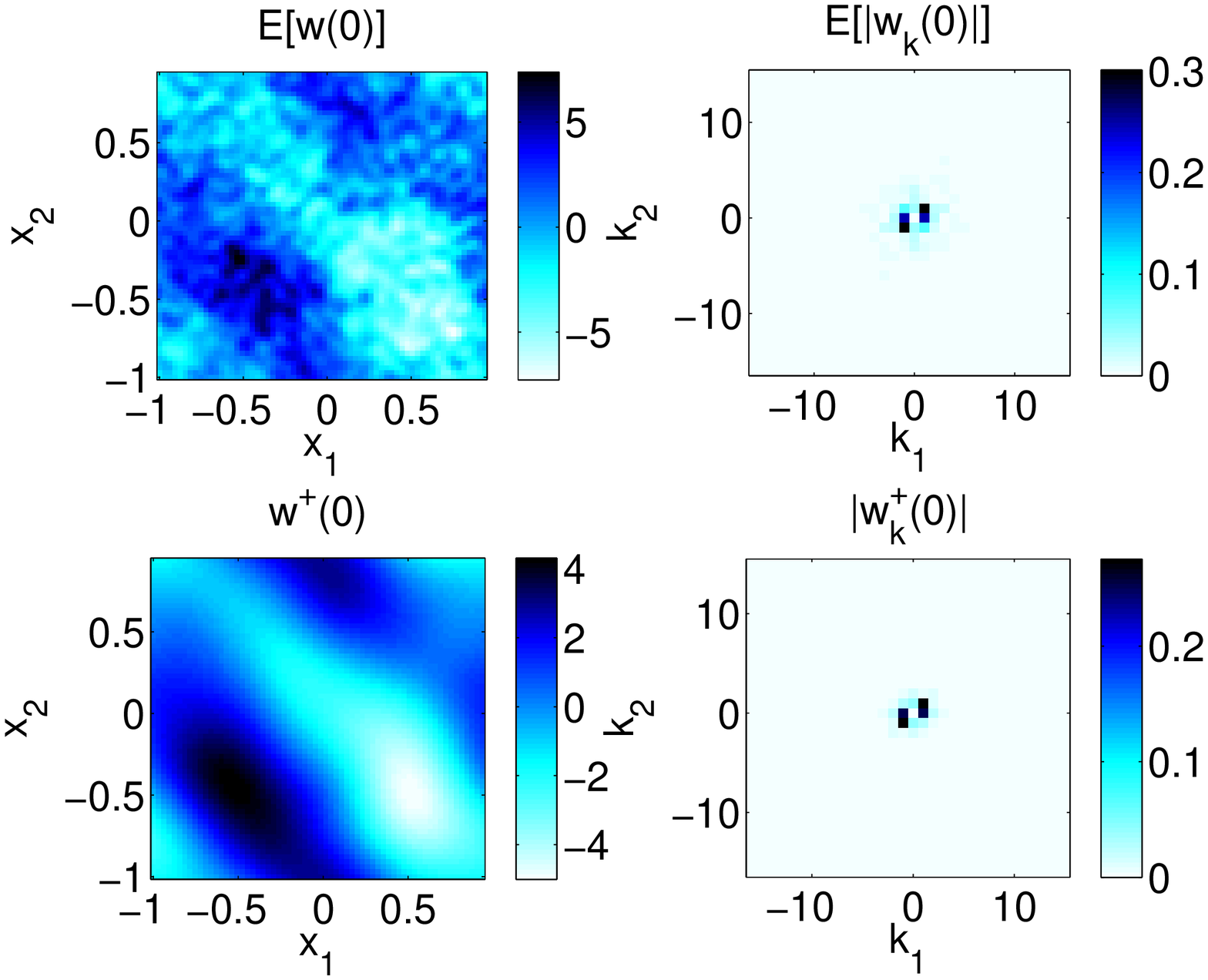}
\caption{Low Reynolds number, stationary solution regime ($\nu=0.1$).  
The vorticity, $w(0)$ (left) of the smoothing distribution at $t=0$, 
and its Fourier coefficients (right), are presented for  
$T=10h=2$.  The top and bottom rows are
the MCMC sample mean and the truth.  The MAP estimator
is not distinguishable from the mean by eye and so is not displayed.  
The prior mean is taken as a
draw from the prior, and hence is not as smooth as the initial
condition.  It is the influence of the prior which makes the MAP
estimator and mean rough, although structurally the same as 
the truth (the solution operator is smoothing, so these fluctuations
are immediately smoothed out - see Fig. \ref{stationary2}).}
\label{stationary1}
\end{figure*}

\begin{figure*}
\includegraphics[width=1\textwidth]{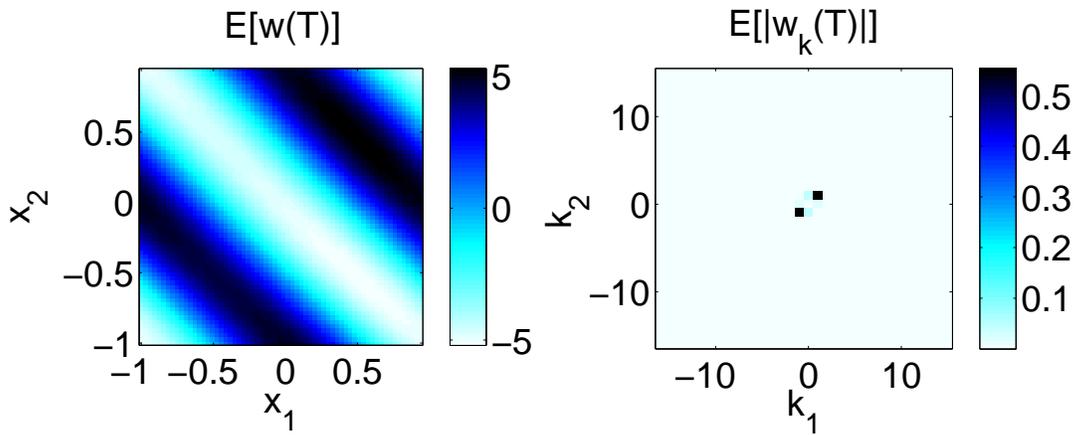}
\caption{Low Reynolds number, stationary solution regime ($\nu=0.1$).  
The vorticity, $w(T)$ (left) of the filtering distribution at $t=T$, 
and its Fourier coefficients (right), are presented for  
$T=10h=2$.   Only the MCMC sample mean is shown, since 
the solutions have been
smoothed out and the difference between the MAP, mean, and truth 
is imperceptible.}
\label{stationary2}
\end{figure*}

\begin{figure*}
\includegraphics[width=1\textwidth]{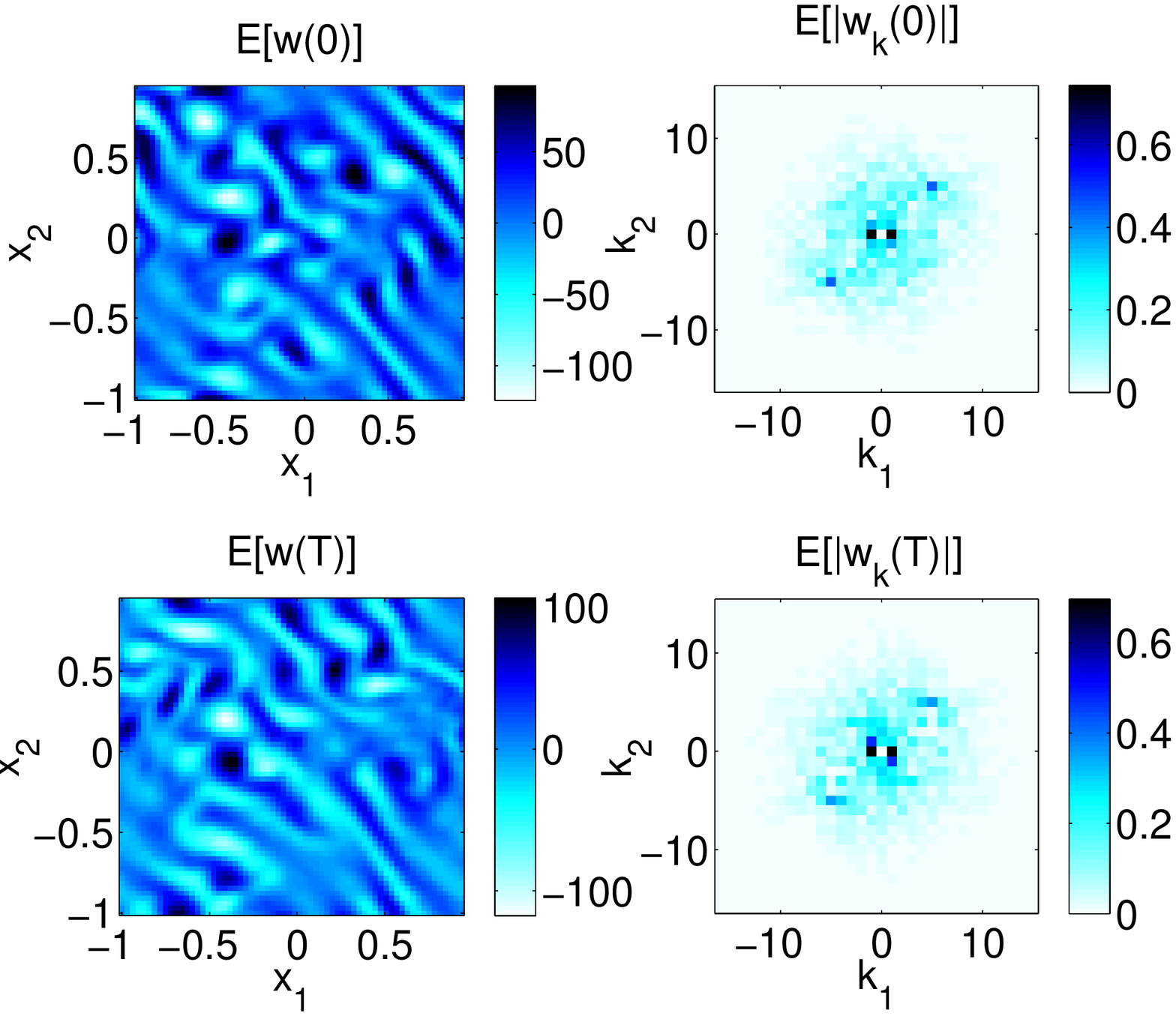}
\caption{
The MCMC mean, as in Fig. \ref{stationary1} for high Reynolds
  number, strongly chaotic solution regime. $\nu=0.01,
  T=10h=0.2, t=0$ (top) and $t=T$(bottom).}
\label{turb1}
\end{figure*}

\begin{figure*}
\includegraphics[width=1\textwidth]{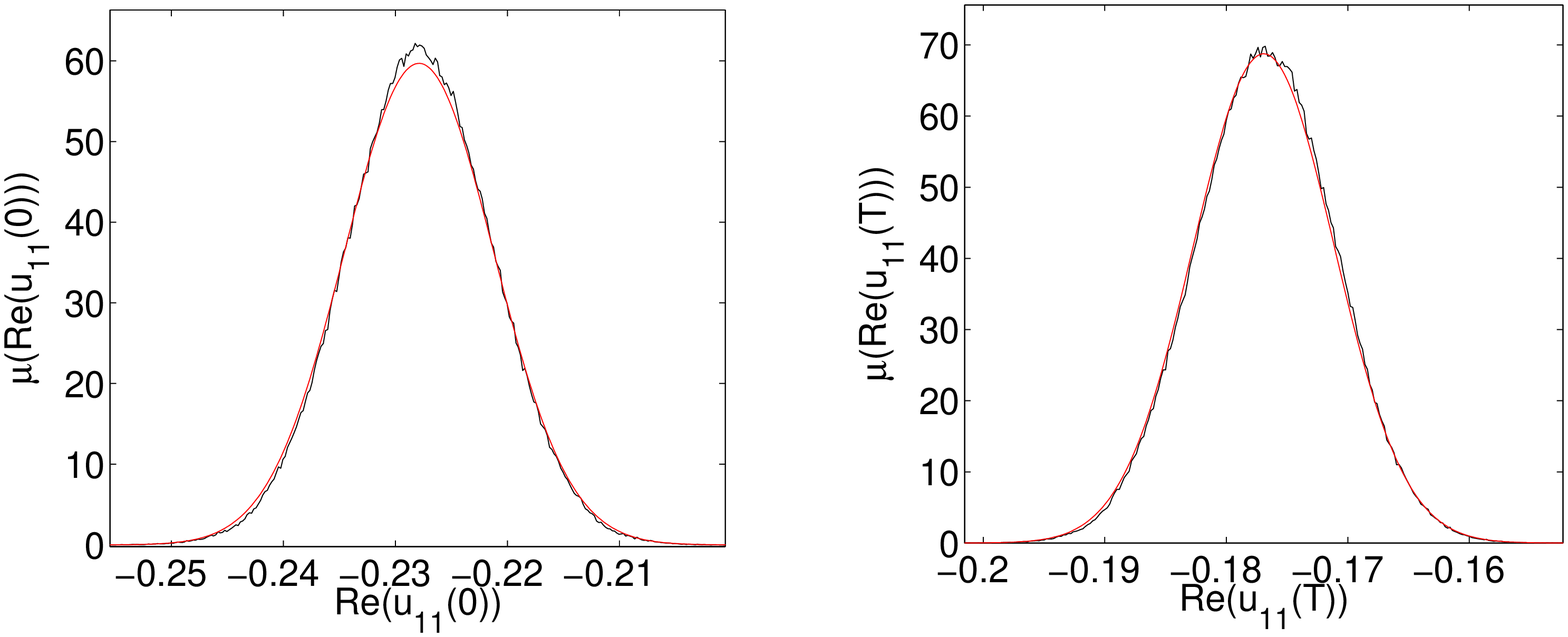}
\includegraphics[width=1\textwidth]{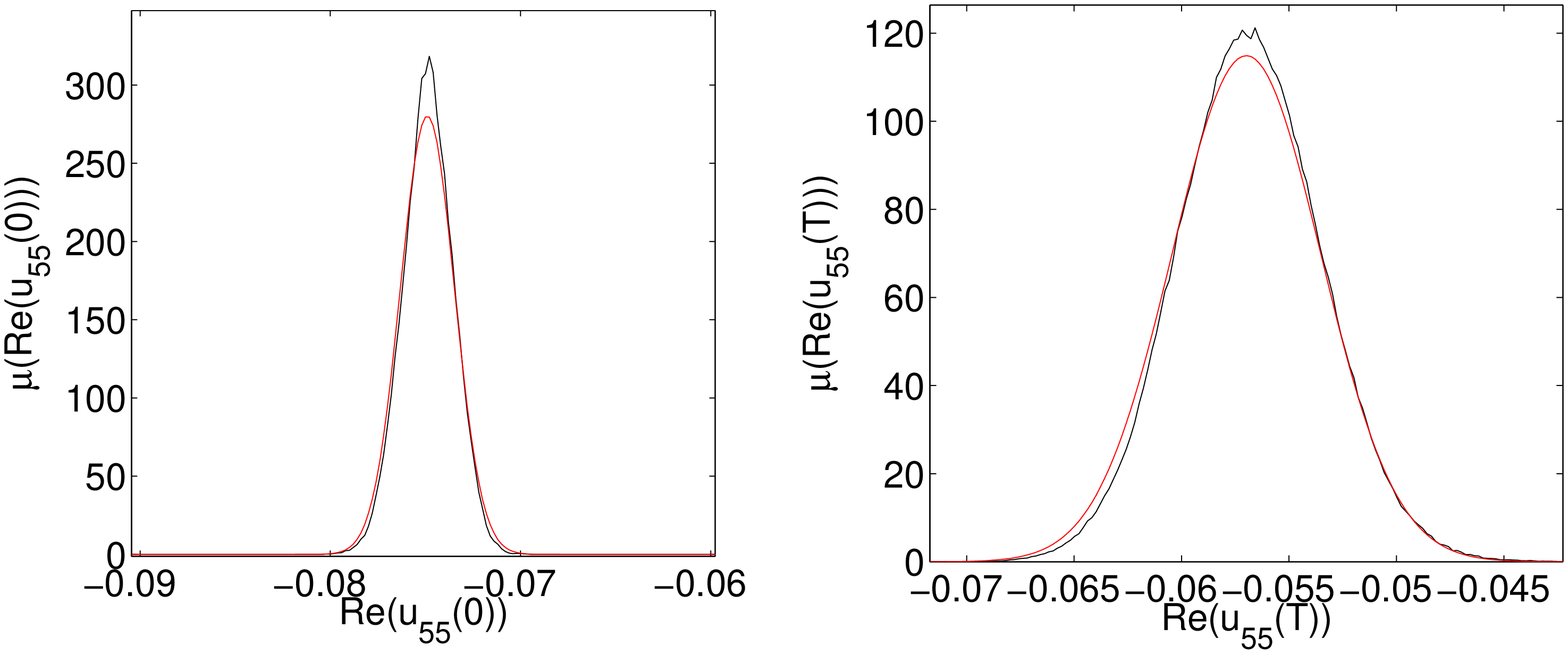}
\caption{Same as Fig. \ref{stationaryv}, except for strongly chaotic regime,
  $\nu=0.01, T=0.2$, and $h=0.02$. The top is mode $u_{1,1}$ and the
  bottom shows mode $u_{5,5}$.}
\label{turbv}
\end{figure*}

Small perturbations in the directions of maximal growth
of the dynamics grow substantially over the larger 
times between observations we look at, 
while over the shorter times the dynamics remain well
approximated by the linearization. 
See the right panel 
of Figure \ref{spectrum} for an example of the local maximal growth 
of perturbations. 
Figure \ref{turb1}
shows the initial and final time profiles of the mean
as in Figures \ref{stationary1} and \ref{stationary2}. 
Now that the solutions themselves are more rough, it is not possible
to notice the influence of the prior mean
at $t=0$, and the profiles of the truth and MAP are indistinguishable 
from the mean throughout
the interval of time.  The situation in this regime is
significantly different from the
situation close to a stationary solution, 
primarily because the dimension of the
attractor is very large and the dynamics on it are very
unpredictable.  
Notice in Figure \ref{turbv} (top) that the uncertainty in $u_{11}$
now barely decreases as we pass from initial
time $t=0$ to final time $t=T$.  
Indeed for moderately high modes, 
the uncertainty increases
(see \ref{turbv} (bottom) for the distribution of 
$u_{55}$).

Table \ref{turbh1} 
presents data for increasing $h=0.02, 0.1, 0.2$, with 
$T=0.2$ fixed.  Table \ref{turbh10_6} 
shows data for increasing $h=0.2, 0.5$ with $T=1$ fixed.
Notable trends, in addition to those mentioned at the
start of the section, are:
(i) when computable, the variance of the 4DVAR 
smoothing distribution has smaller error at $t=0$
than at $t=T$;
(ii) the 4DVAR smoothing distribution error with respect
  to the variance cannot be computed accurately for $T=1$
because of accumulated error for long times in the aproximation
of the adjoint of the forward operator by the discretization of the
analytical adjoint;
(iii) the error of 4DVAR with respect to the mean at
$t=0$ for $h \leq 0.1$
is below the threshold of accuracy of MCMC;  
(iv) the error in the variance for the FDF algorithm
is very large because the $\cQ$ is an order of
magnitude larger than $\Gamma$;
(v) the FDF algorithm is consistent in recovering the mean for
  increasing $h$, while the other algorithms deteriorate;
(vi) the error of FDF with 
respect to the variance decreases with
  increasing $h$;
(vii) for $h=0.5$ and
$T=1$ the FDF performs best and 
these desirable properties of the FDF variant on 3DVAR
are associated with stability and will be discussed in the
next section;
(viii) for increasing $h$, the error in the mean of 
LRExKF increases first when $h=0.1$ and $T=0.2$ and becomes
close to the error in the variance which can be
explained  by the bias induced by neglecting the 
next order of the expansion of the dynamics;
(ix) the error in LRExKF is substantial when $T=1$ and 
it really majorly fails when $h=0.5$ which is
consistent with the time-scale on which nonlinear 
effects become prominent (see Fig. \ref{spectrum}) and the 
linear approximation would not be expected to be valid.  
The error in the mean is larger, again as expected from the Ito 
correction term.

\begin{table*}
\begin{center}
\begin{tabular}{|c||c|c|c|c|c|} \hline
$h=0.2$       &$e_{mean}$    &$e_{variance}$&$e_{truth}$   &$e_{obs}$     &$e_{map}$      \\ \hline \hline
 MCMC($t=0$)  & 0 & 0 & 0.17177 & 0.819094 & 0.00153443 \\ \hline
 4DVAR($t=0$) & 0.00153523 & 0.00620345 & 0.185876 & 0.740612 & 0 \\ \hline
 MCMC($t=T$)  & 0 & 0 & 0.0164605 & 0.558026 & 5.17207e-05 \\ \hline
 4DVAR($t=T$) & 5.1723e-05 & 0.00459055 & 0.0164618 & 0.558024 & 0 \\ \hline
3DVAR        & 0.138652 & 108.516 & 0.13738 & 0.54585 & 0.138646 \\ \hline
 FDF       & 0.00173093 & 0.423299 & 0.0153513 & 0.558228 & 0.00172455 \\ \hline
 LRExKF       & 6.34566e-05 & 0.00320937 & 0.0164796 & 0.558022 & 2.22202e-05 \\ \hline
EnKF        & 0.00359669 & 0.119076 & 0.0158585 & 0.558032 & 0.00362309 \\ \hline
 truth ($t=0$)& 0.17177 & - & 0 & 0.816333 & 0.156072 \\ \hline
 truth ($t=T$)& 0.0164605 &- & 0 & 0.713754 & 0.0164342 \\ \hline
 \hline \hline
$h=1$        &$e_{mean}$    &$e_{variance}$&$e_{truth}$   &$e_{obs}$     &$e_{map}$      \\ \hline \hline
 MCMC($t=0$)  & 0 & 0 & 0.295424 & 0.791832 & 0.00110927 \\ \hline
 4DVAR($t=0$) & 0.00110969 & 0.00375462 & 0.333225 & 0.748439 & 0 \\ \hline
 MCMC($t=T$)  & 0 & 0 & 0.028831 & 0.662342 & 0.00016539 \\ \hline
 4DVAR($t=T$) & 0.000165408 & 0.00896381 & 0.0287779 & 0.662373 & 0 \\ \hline
 3DVAR        & 0.128956 & 41.6646 & 0.139419 & 0.646462 & 0.128929 \\ \hline
 FDF           & 0.00400194 & 0.458239 & 0.031512 & 0.654203 & 0.00403853 \\ \hline
 LRExKF       & 0.000165666 & 0.00267976 & 0.0287787 & 0.65413 & 1.84537e-05 \\ \hline
EnKF        & 0.00289635 & 0.122461 & 0.0301991 & 0.654205 & 0.00285458 \\ \hline
 truth ($t=0$)& 0.295424 & - & 0 & 0.780891 & 0.27957 \\ \hline
 truth ($t=T$)& 0.028831 & - & 0 & 0.77011 & 0.0287068 \\ \hline
\hline \hline
$h=2$        &$e_{mean}$    &$e_{variance}$&$e_{truth}$   &$e_{obs}$     &$e_{map}$      \\ \hline \hline
 MCMC($t=0$)  & 0 & 0 & 0.32043 & 0.747756 & 0.000965003 \\ \hline
 4DVAR($t=0$) & 0.000965294 & 0.00384239 & 0.357404 & 0.633977 & 0 \\ \hline
 MCMC($t=T$)  & 0 & 0 & 0.03871 & 0.68846 & 0.000208273 \\ \hline
 4DVAR($t=T$) & 0.000208299 & 0.00250571 & 0.0386606 & 0.68846 & 0 \\ \hline
 3DVAR        & 0.105535 & 35.9905 & 0.108918 & 0.684345 & 0.10548 \\ \hline
 FDF          & 0.00177839 & 0.475338 & 0.0387006 & 0.688477 & 0.00173164 \\ \hline
 LRExKF       & 0.0002106 & 0.00272041 & 0.0386602 & 0.68846 & 2.991e-06 \\ \hline
EnKF      & 0.00319756 & 0.106976 & 0.0385305 & 0.688464 & 0.00312047 \\ \hline
 truth ($t=0$)& 0.32043 & - & 0 & 0.771936 & 0.299957 \\ \hline
 truth ($t=T$)& 0.03871 & - & 0 & 0.688664 & 0.038578 \\ \hline
\end{tabular} 
\end{center}
\caption{Stationary state regime, $\nu=0.1$, $T=2$, with $h=0.2$ (top
  table), $h=1$ (middle), and $h=2$ (bottom). 
The first, third, fourth and fifth columns are 
the norm difference, $e=||M-m||/||M||$, 
where $M$ is the mean of the posterior distribution (MCMC),
the truth, the observation, or the MAP estimator and $m$ is the 
mean obtained from the various methods.  
The second column is the norm difference, 
$e=||{\rm var}[u]-{\rm var}[U]||/||{\rm var}[u]||$
where ${\rm var}$ indicates the variance, $u$ is sampled from
the posterior (via MCMC), 
and $U$ is the approximate state obtained from the various methods.}
\label{stationaryh1}
\end{table*}

\section{Filter Stability}
\label{stabilitysec}

Many of the accuracy results for the filters described
in the previous section are degraded if, as is common
practice in applied scenarios, modifications are made
to ensure that the algorithms remain stable over longer
time-intervals; that is if some form of variance inflation 
is performed to keep the algorithm close to the true signal, 
or to prevent it from suffering filter divergence (see
\cite{jazwinski1970stochastic}, \cite{fisher2005equivalence},
\cite{evensen2009data}, and references therein).
In this section we describe some of the mathematics which
underlies stabilization, describe numerical results
illustrating it, and investigate its effect on filter
accuracy. {\em The basic conclusion of this
section is that stabilization via
variance inflation enables algorithms to be run for longer
time windows before diverging, but may cause poorer
accuracy in both the mean (before divergence) and the variance predictions.}
Again, we make no claims of optimal implementation of these
filters, but rather aim to describe the mechanism of 
stabilization and the common effect, in general, 
as measured by ability to reproduce the gold standard posterior 
distribution.

We define stability in this context to mean
that the distance between the truth and the estimated mean 
remains small. As we will demonstrate, whether or not 
this distance remains small depends on whether
the observations made are sufficient to control any 
instabilities inherent in the model dynamics.
To understand this issue it is instructive to consider the
3DVAR, FDF and LRExKF filters, all of which use a prediction
step \eqref{forward2} which updates the mean
using $m_j=\Psi({\hat m}_{j-1})$. When combined
with the data incorporation step \eqref{eq:mean}
we get an update equation of the form
\begin{linenomath*}
\begin{equation}
\hat{m}_{j+1} = (I-K_j) \Psi(\hat{m}_j) + K_j y_{j+1}, 
\label{markov_est}
\end{equation}
\end{linenomath*}
where $K_j=\bigl(\cC_j^{-1}+\Gamma^{-1}\bigr)^{-1}\Gamma^{-1}$
is the Kalman gain matrix.
If we assume that the data is derived from a true signal
$u_j^\dagger$ satisfying $u_{j+1}^\dagger=\Psi(u_j^\dagger)$
and that 
$$y_{j+1}=u_{j+1}^\dagger+\eta_j=\Psi(u_j^\dagger)+\eta_j,$$ 
where the $\eta_j$
denote the observation errors,
then we see that \eqref{markov_est} has the form
\begin{linenomath*}
\begin{equation}
{\hat m}_{j+1} = (I-K_j) \Psi({\hat m}_j) + K_j\Psi(u_j^\dagger) 
+K_j \eta_{j+1}.
\label{markov2}
\end{equation}
\end{linenomath*}
If the observational noise is assumed to be consistent
with the model used for the assimilation, then
$\eta_j \sim \cN(0,\Gamma)$ are i.i.d. random variables and 
we note that \eqref{markov2} is an inhomogenous Markov chain.

Note that
\begin{linenomath*}
\begin{equation}
u_{j+1}^{\dagger} = (I-K_j) \Psi(u_j^{\dagger}) + K_j\Psi(u_j^\dagger) 
\label{markov3}
\end{equation}
\end{linenomath*}
so that defining the error 
$e_j:=\hat{m}_{j} - u_{j}^\dagger$ 
and subtracting \eqref{markov3} from \eqref{markov2} 
we obtain the equation
$$e_{j+1} \approx  (I-K_j) D_je_j + K_j \eta_{j+1}$$
where $D_j=D\Psi(u_j^\dagger).$
The stability of the filter will be governed by families of
products of the form 
\begin{linenomath*}
\begin{equation}
\nonumber
\Pi_{j=0}^{k-1}\bigl((I-K_j) D_j\bigr),\quad k=1,\dots, J.
\end{equation}
\end{linenomath*}
We observe that $I-K_j$ will act to induce stability, as it
has norm less than one in appropriate spaces; $D_j$,
however, will induce some instability whenever the dynamics
themeslves contain unstable growing modes.
The balance between these effects -- stabilization through
observation and instability in the dynamics -- determines
whether the overall algorithm is stable.

The operator $K_j$ weights the relative importance
of the model and the observations, according to
covariance information.  
Therefore, this weighting must effectively 
stabilize the growing directions in the dynamics.
Note that increasing $\cC_j$ 
-- {\em variance inflation} --
has the effect of moving $K_j$ towards the identity, 
so the mathematical mechanism of controlling 
the instability mechanism by variance inflation
is elucidated by the discussion above. 
In particular, when the assimilation is proceeding
in a stable fashion, the modes in which growing
directions have support typically 
overestimate the variance to induce this stability.
In unstable cases, there are at least some times at which
some modes in which growing directions have support
{\it underestimate} the variance, leading to instability
of the filter.
It is always the case that the onset of instability occurs
when the distance from the estimated mean to the 
truth persistently exceeds the estimated standard deviation.  
In \cite{stuartlaw_nlin} we provide the
mathematical details and rigorous proofs which underpin the
preceding discussion.

In the following, two observations concerning 
the size of the error
are particularly instructive. Firstly,
using the distribution assumed on the $\eta_j$, the following
lower bound on the error is immediate\footnote{Here
$\bbE$ denotes expectation with respect to the random variables
$\eta_j$.}:
\begin{linenomath*}
\begin{equation}
\bbE \|e_{j+1}\|^2 \ge \bbE\|K_j\eta_{j+1}\|^2=
{\rm tr}\bigl(K_j\Gamma K_j^*\bigr).
\label{eq:lower}
\end{equation}
\end{linenomath*}
This implies that the average scale of the error of the
filter, with respect to the truth, is set by the scale
of the observation error,  and shows that the choice of
the covariance updates, and hence the Kalman gain $K_j$, will affect
the exact size of the average error, on this scale.
The second observation follows from considering the
trivial ``filter'' obtained by setting $K_j \equiv I,$
which corresponds to simply setting ${\hat m}_{j}=y_j$
so that all weight is placed on the observations.
In this case the average error is equal to 
\begin{linenomath*}
\begin{equation}
\label{eq:upper}
\bbE\|e_{j+1}\|^2=
\bbE\|\eta_{j+1}\|^2={\rm tr}(\Gamma).
\end{equation}
\end{linenomath*}
As we would hope that incorporation of the model itself
improves errors we view \eqref{eq:upper} as providing
an upper bound on any reasonable filter and we will consider
the filter ``unstable'' if the squared error from the truth
exceeds 
${\rm tr}(\Gamma)$. Thus we use \eqref{eq:upper}
and \eqref{eq:lower} as guiding upper and lower bounds
when studying the errors in the filter means in what follows.

In cases where our basic algorithm is unstable in the
sense just defined we will also implement a stabilized
algorithm, by adopting the commonly used
practice of variance inflation. The 
discussion above
demonstrates how this acts to induce stability
by causing the $K_j$ to move closer to the identity. 
For 3DVAR this is achieved by taking 
the original $\cC_0$ and redefining it via
the transformation 
$\cC_0 \rightarrow \frac{1}{\epsilon}\cC_0$.  
In all the numerical computations presented in this paper
which concern the stabilized version of 3DVAR
we take $\epsilon=0.01.$
The FDF(b) algorithm remains stable since it already has
an inflated variance via the model error term.
For 
LRExKF we achieve variance inflation
by replacing the perturbation
term of Equation \ref{covariance2} with
$(I-VV^*)\tilde{\cC}_j(I-VV^*)$, where
$\tilde{\cC}_j$ is the covariance arising 
from FDF(b).
Finally we discuss stabilization of the EnKF. 
This is achieved by taking 
the original $\cC_j$'s given
by \eqref{enkf:cov} 
and redefining them via
the transformations 
$\cC_0 \rightarrow
\frac{1}{\epsilon}\cC_0$, and 
$\cC_j \rightarrow
(1+\varepsilon_i)\cC_j+\varepsilon_p \cC_0$ with
$\epsilon=10^{-4}, \varepsilon_i=0.1, \varepsilon_p=0.01$.
The parameter $\epsilon$ prevents initial divergence,
$\varepsilon_i$ maintains stability with direct incremental 
inflation and $\varepsilon_p$ provides rank correction.  
This is only one option out of a wide array of possible 
hueristically derived such transformations.
For example, rank correction is often performed by 
some form of localization which preserves trace and eliminates
long-range correlations, while our
rank correction preserves long-range correlations and 
provides trace inflation.
The point here is that our transformation captures the 
essential mechanism of stabilization by inflation which, 
again, is our objective. 

We denote the
stabilized versions of 3DVAR, LRExKF, and EnKF by
[3DVAR], [LRExKF], and [EnKF].  Because FDF itself always
remains stable we do not show results for a stabilized
version of this algorithm.  Note that we
use ensembles in EnKF of equal size to the number of 
approximate eigenvectors in LRExKF, in order to ensure 
comparable work.  This is always 100, except for large $h$,
when some of the largest 100 eigenvalues are too close to zero
to maintain accuracy, 
and so fewer eigenvectors are used in LRExKF in these cases.  
Also, note again that we are looking for general features
across methods and are not aiming to optimize the inflation 
procedure for any particular method.

\begin{figure*}
\includegraphics[width=.45\textwidth]{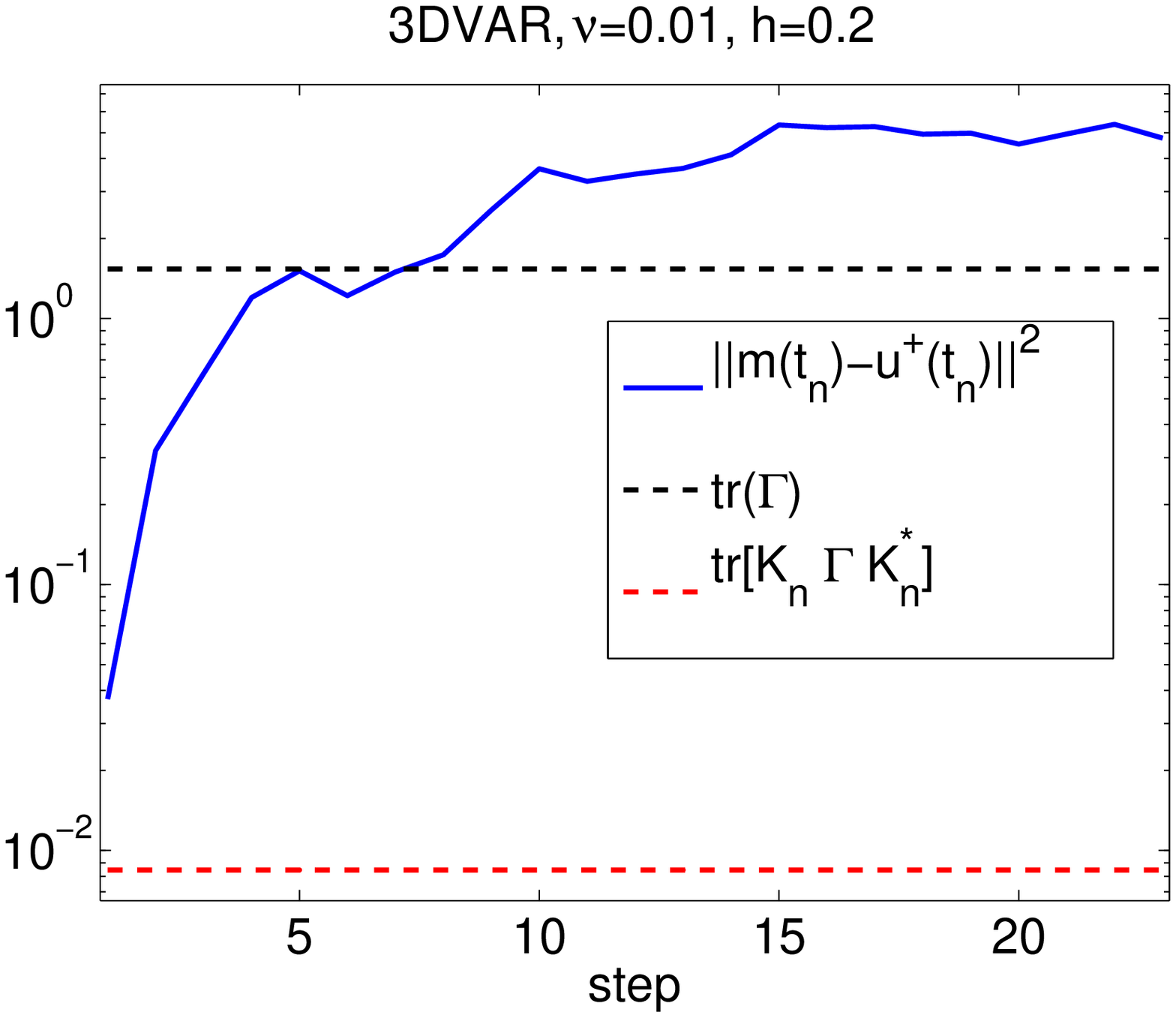}
\includegraphics[width=.45\textwidth]{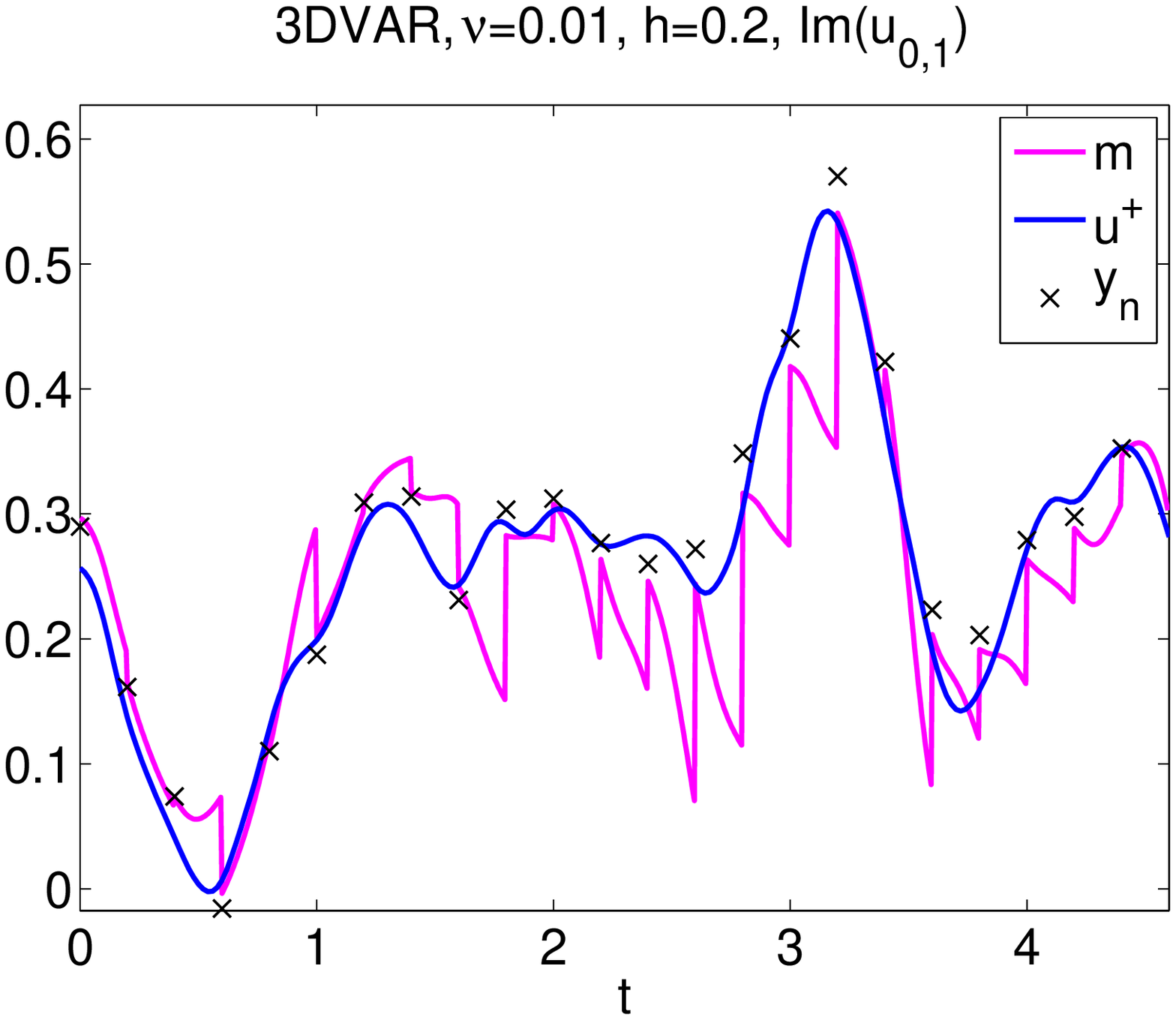}
\includegraphics[width=.45\textwidth]{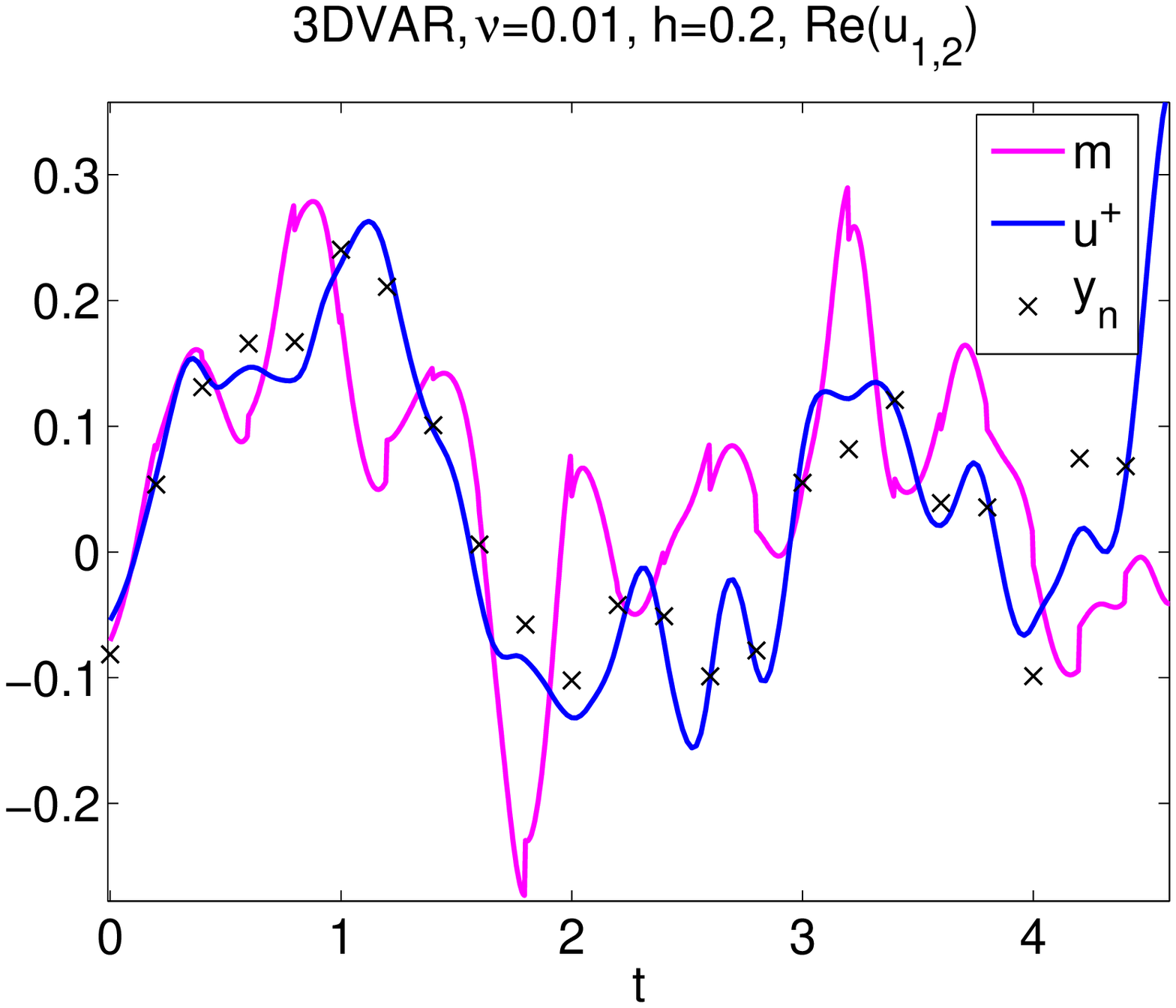}
\includegraphics[width=.45\textwidth]{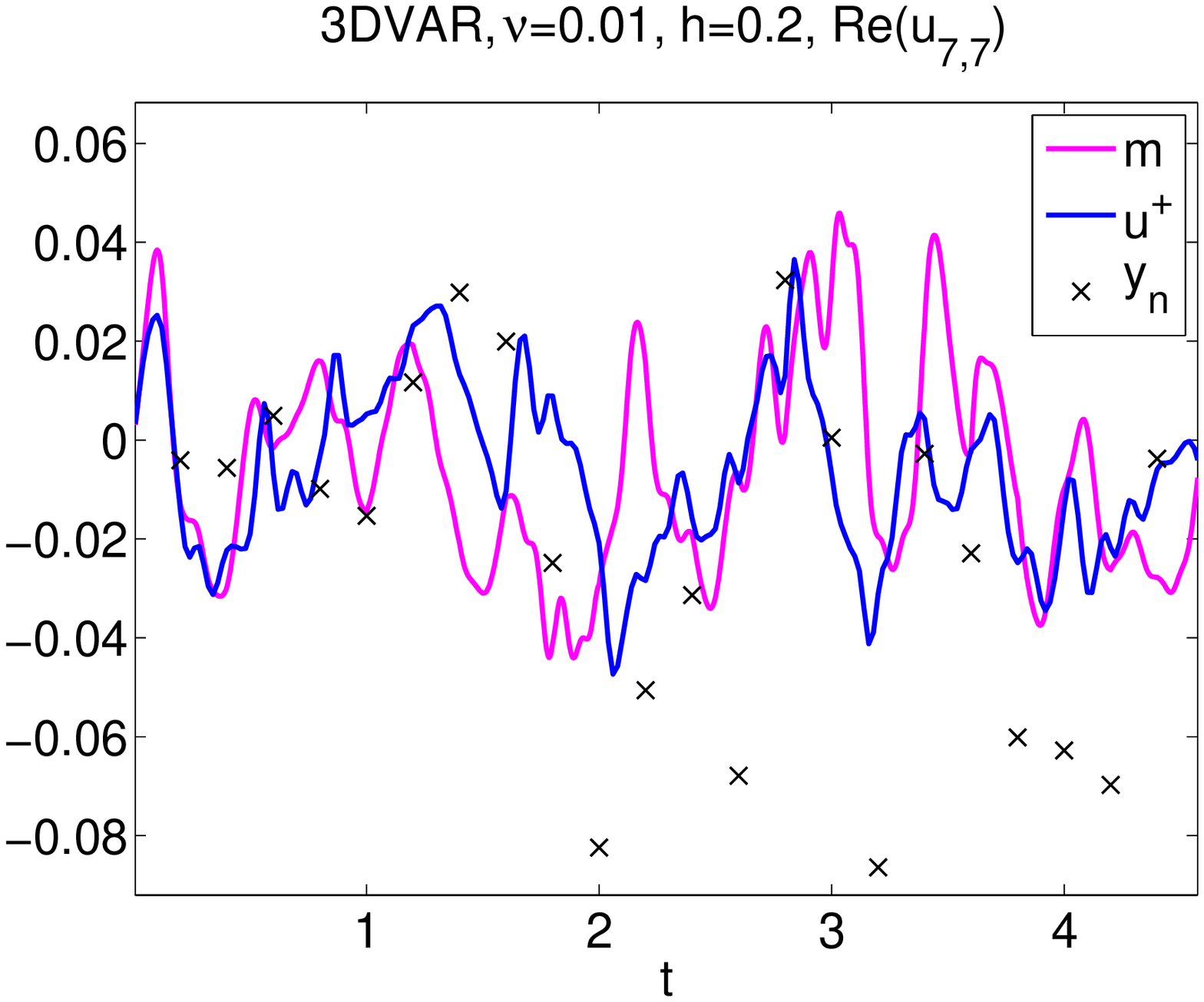}
\caption{Example of an unstable trajectory for 3DVAR with $\nu=0.01,
  h=0.2$.  The top left plot shows the norm-squared error between the
estimated mean, $m(t_n)=\hat{m}_n$, and the truth, $u^{\dagger}(t_n)$, 
in comparison to the preferred upper
bound (i.e. the total observation error ${\rm tr} (\Gamma)$,
\eqref{eq:upper}) and the lower bound 
${\rm tr} [K_n \Gamma K_n^*]$ \eqref{eq:lower}.  The other three plots show the 
estimator, $m(t)$, together with the truth, $u^{\dagger}(t)$, and the
observations, $y_n$ for a few individual modes.}
\label{var_unstable}
\end{figure*}

\begin{figure*}
\includegraphics[width=.45\textwidth]{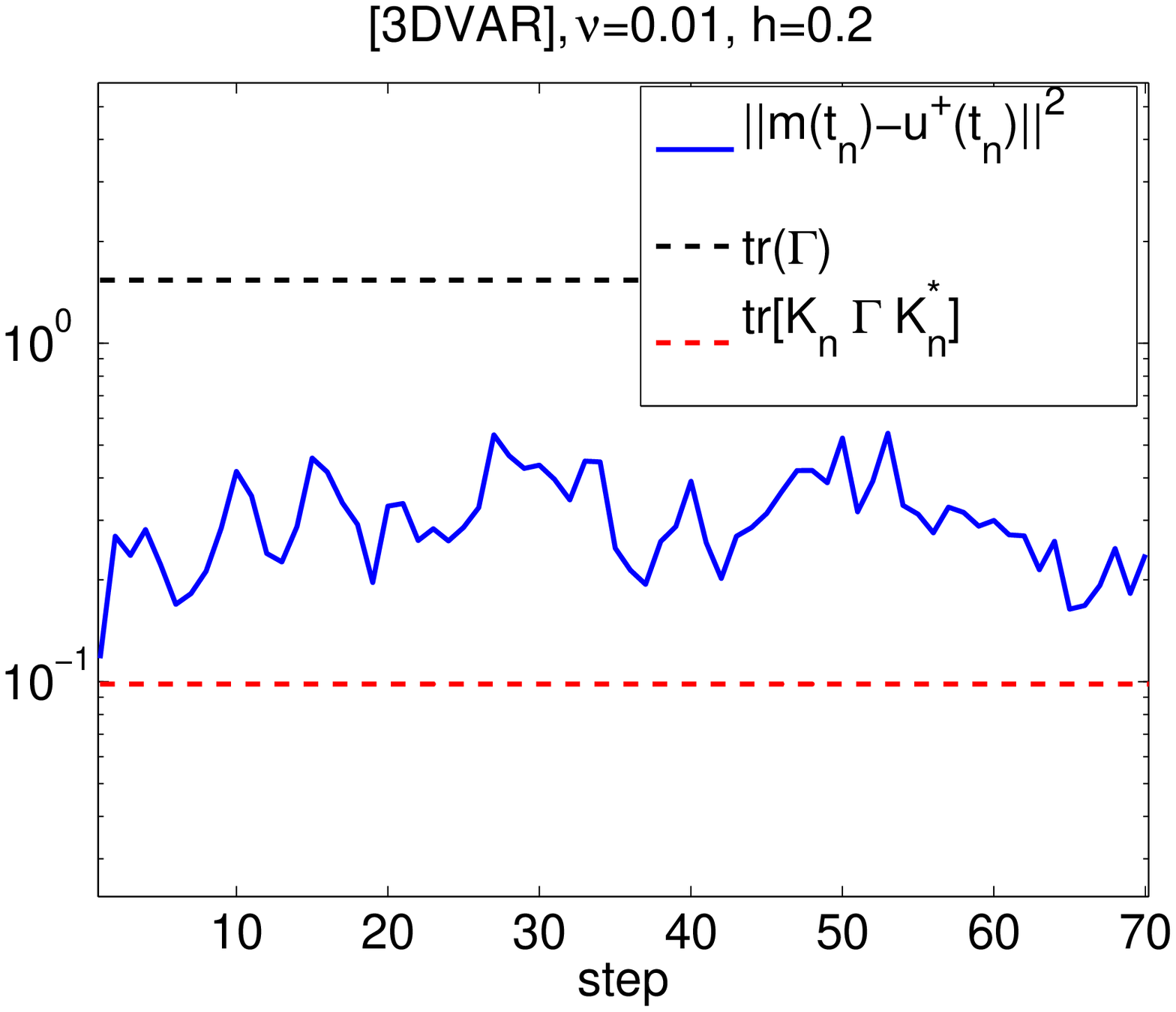}
\includegraphics[width=.45\textwidth]{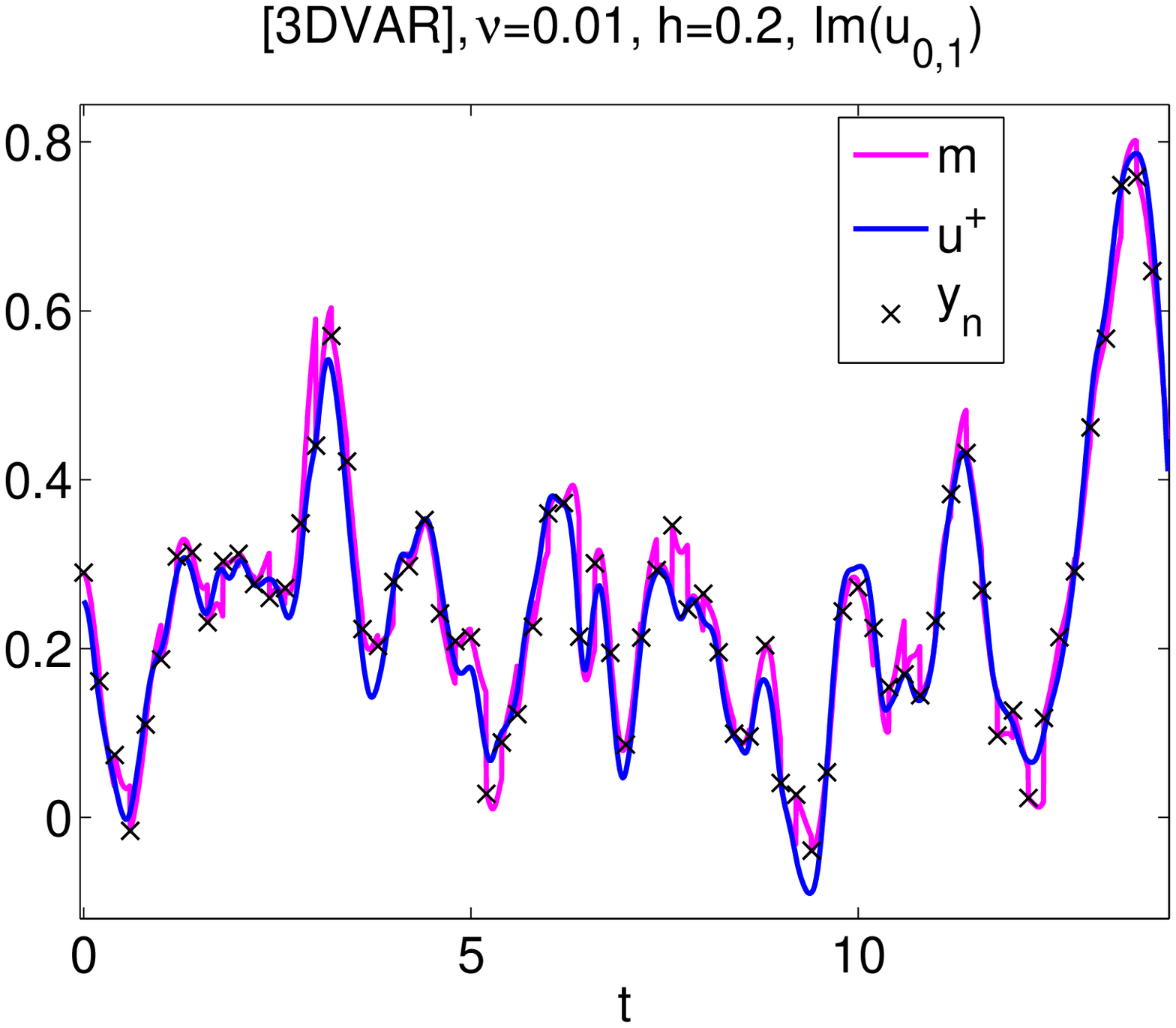}
\includegraphics[width=.45\textwidth]{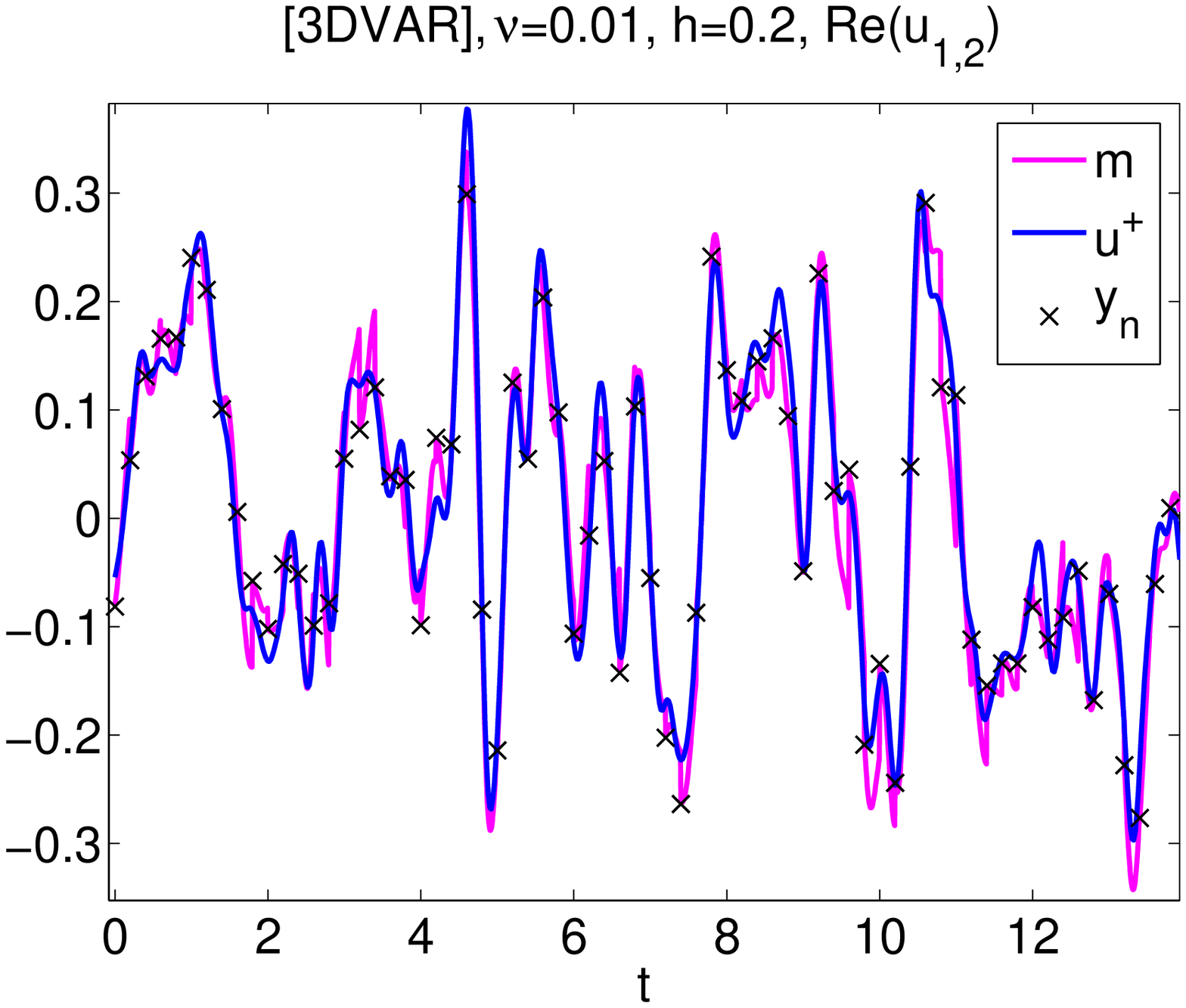}
\includegraphics[width=.45\textwidth]{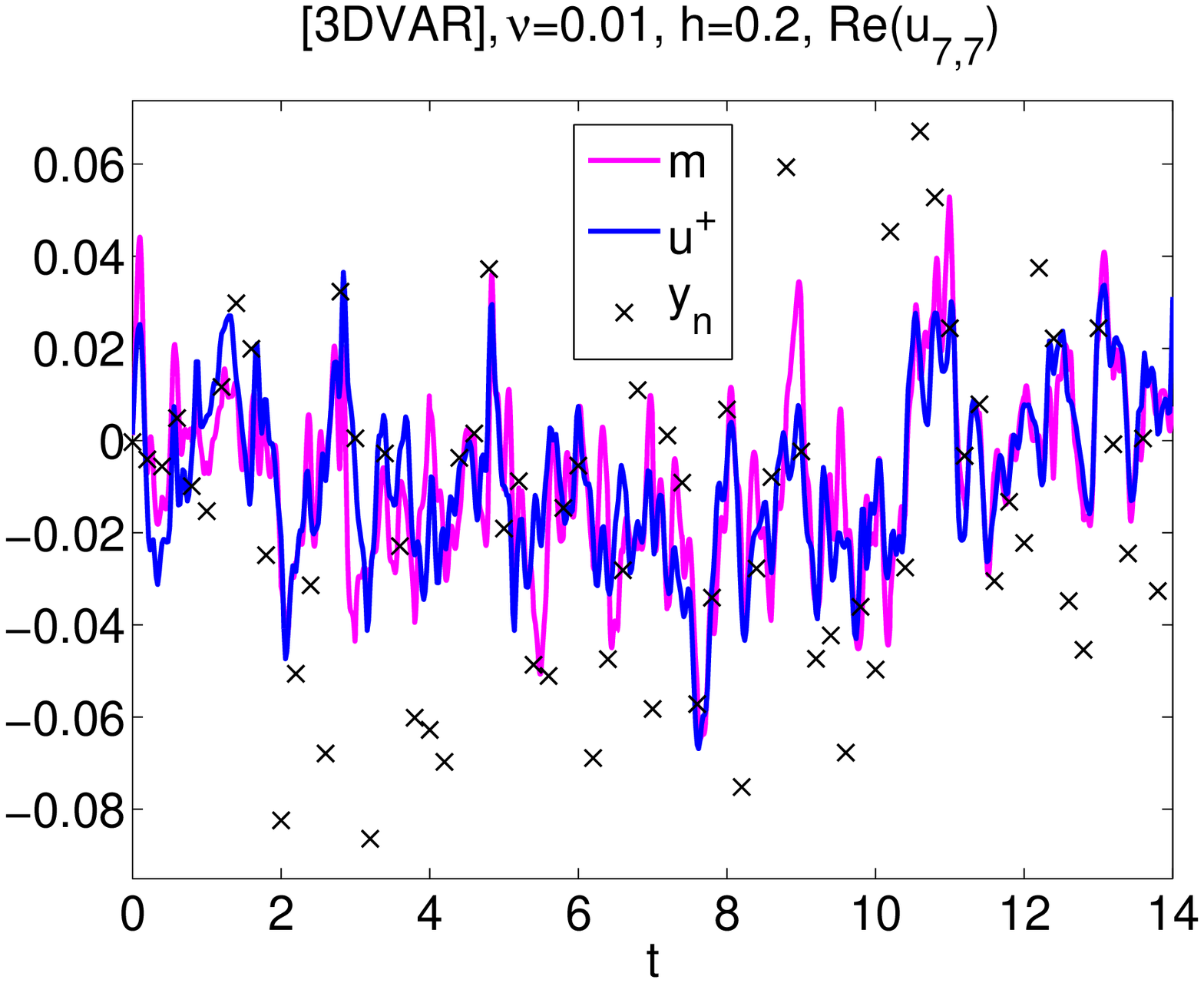}
\caption{Example of a variance-inflated stablilized trajectory ($\cC_0
  \rightarrow \frac{1}{\epsilon} \cC_0$) for
  [3DVAR] with the same external parameters as in
  Fig. \ref{var_unstable}.
Panels are the same as in Fig. \ref{var_unstable}.}
\label{var_stable}
\end{figure*}

\begin{figure*}
\includegraphics[width=.45\textwidth]{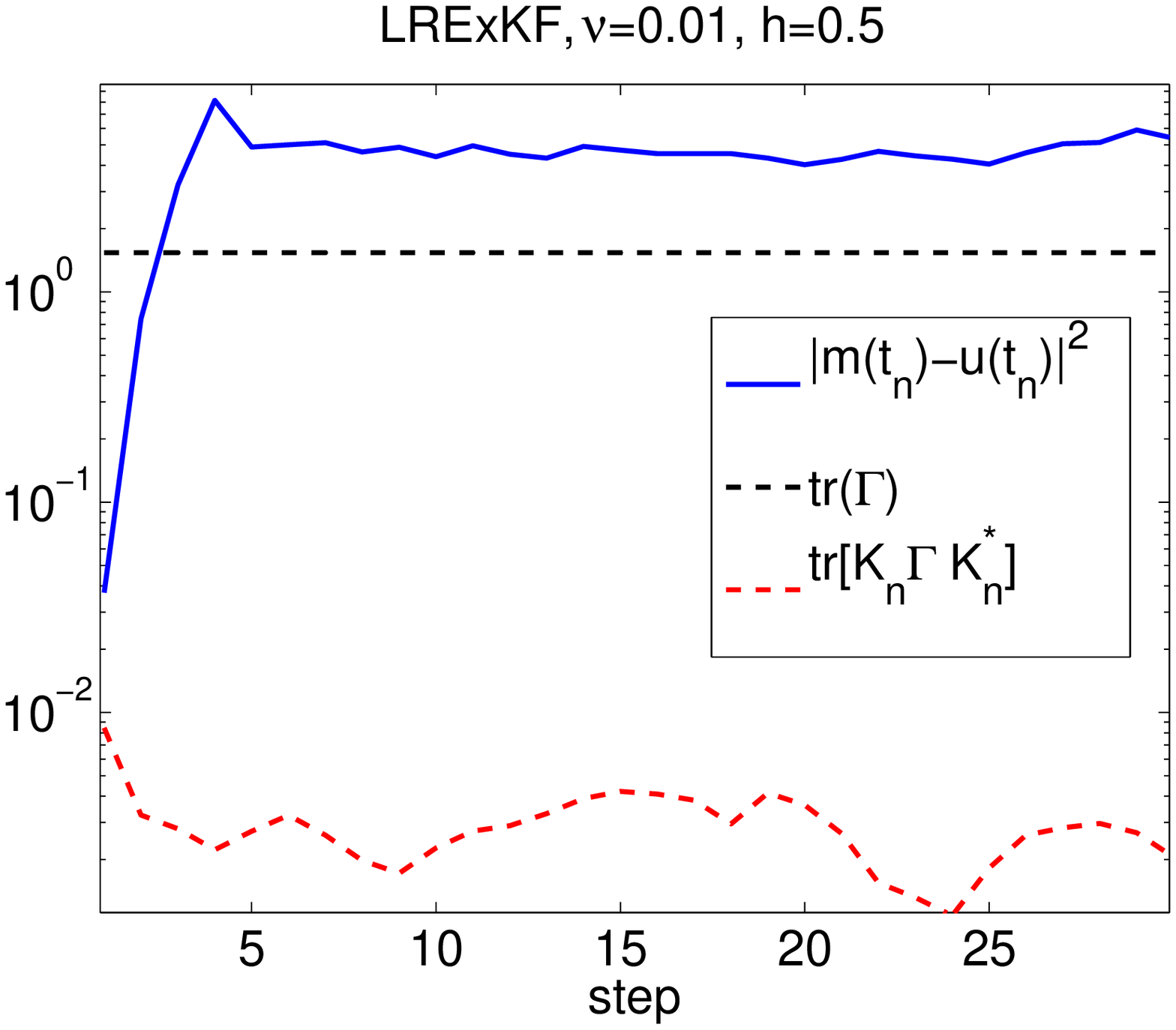}
\includegraphics[width=.45\textwidth]{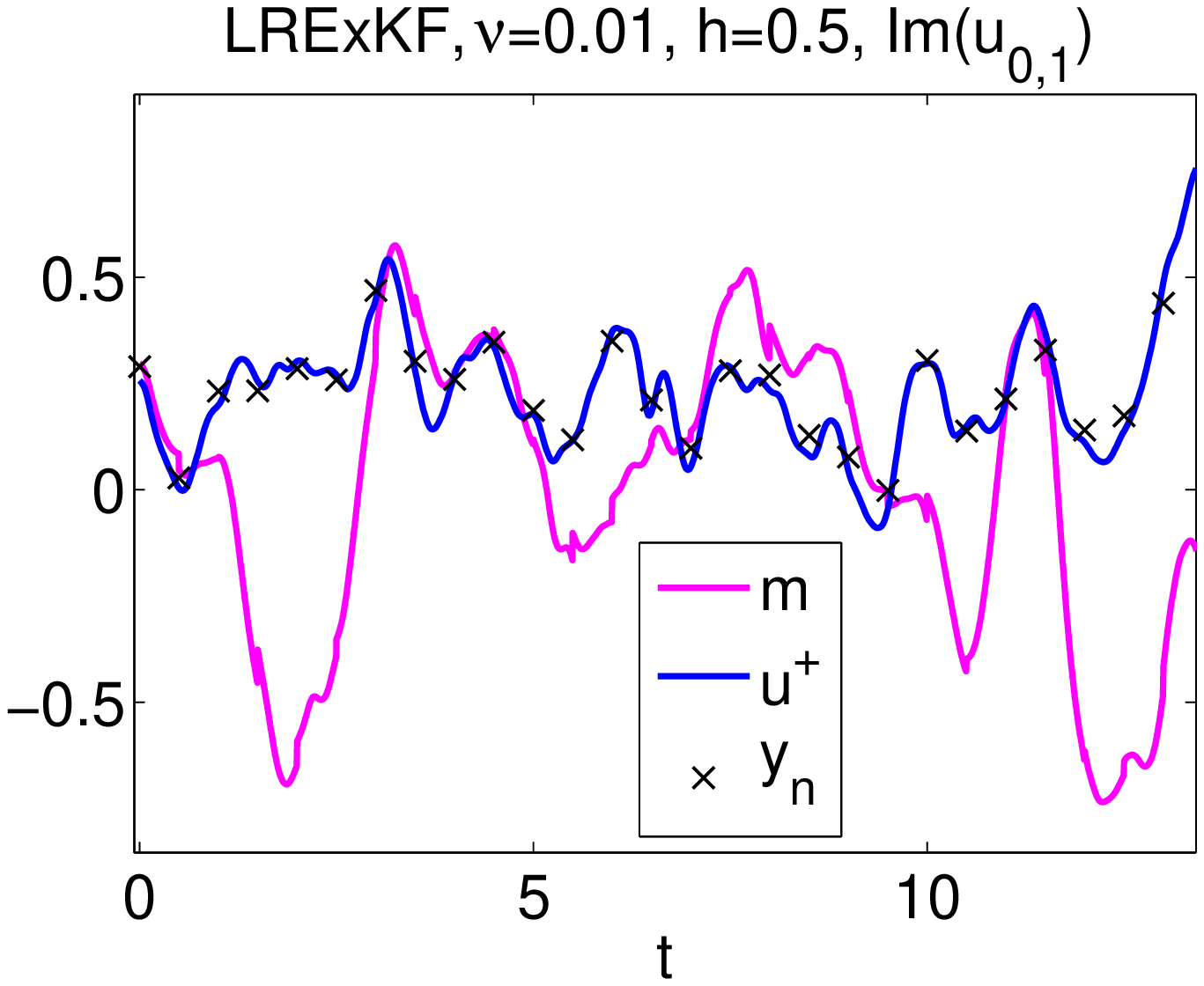}
\includegraphics[width=.45\textwidth]{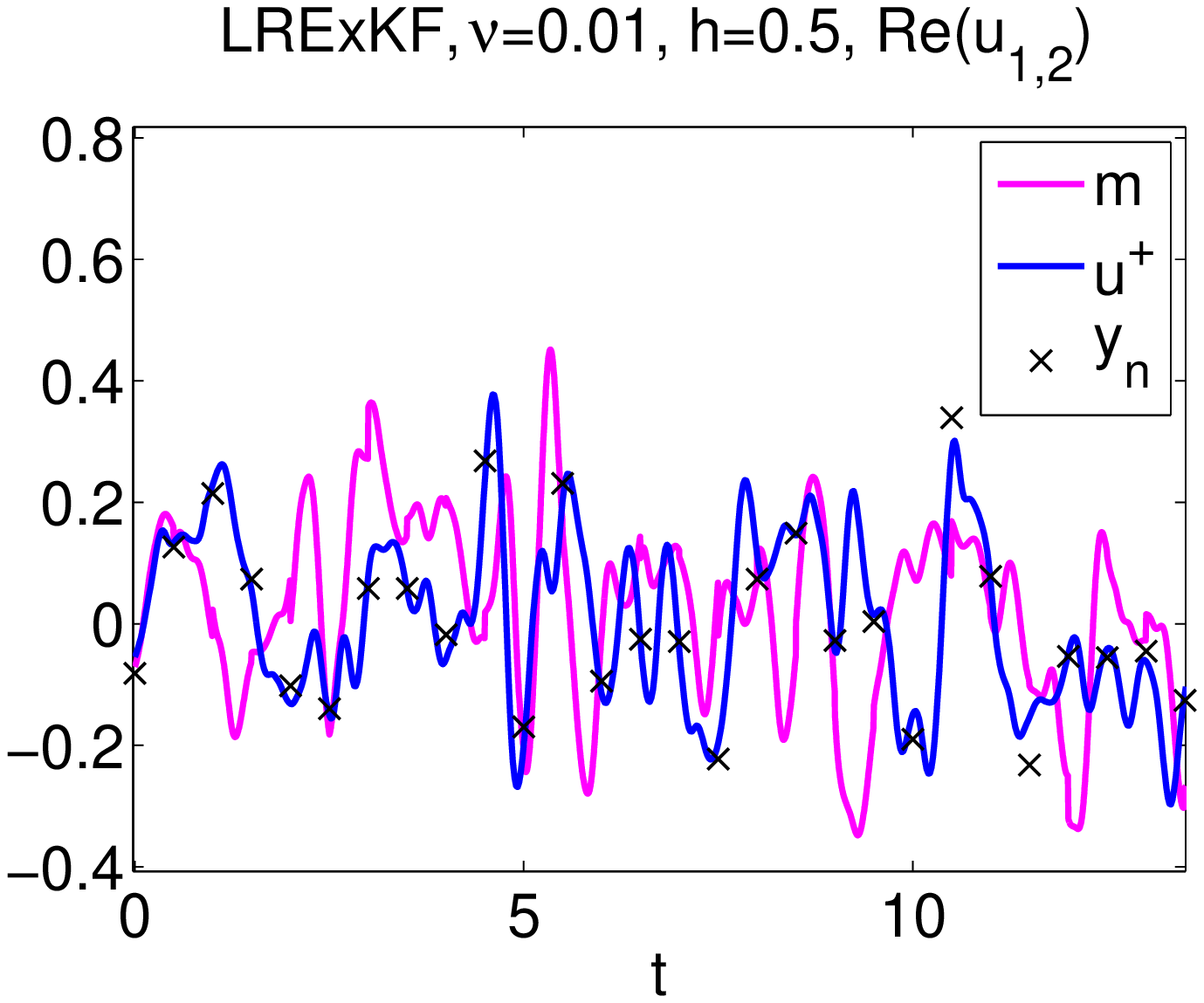}
\includegraphics[width=.45\textwidth]{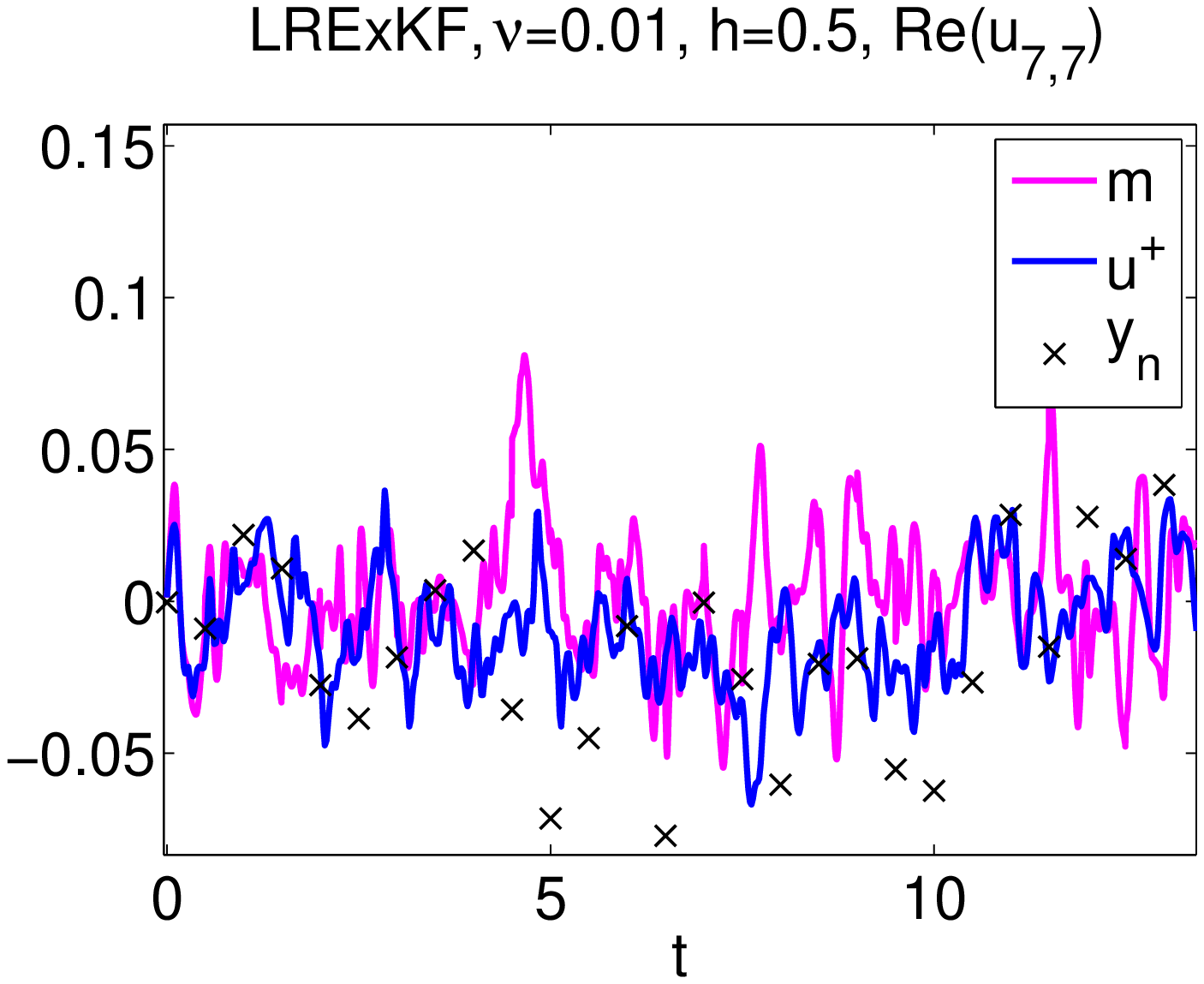}
\caption{Example of an unstable trajectory for LRExKF with $\nu=0.01,
  h=0.5$. Panels are the same as in Fig. \ref{var_unstable}.} 
\label{fdf_unstable}
\end{figure*}

\begin{figure*}
\includegraphics[width=.45\textwidth]{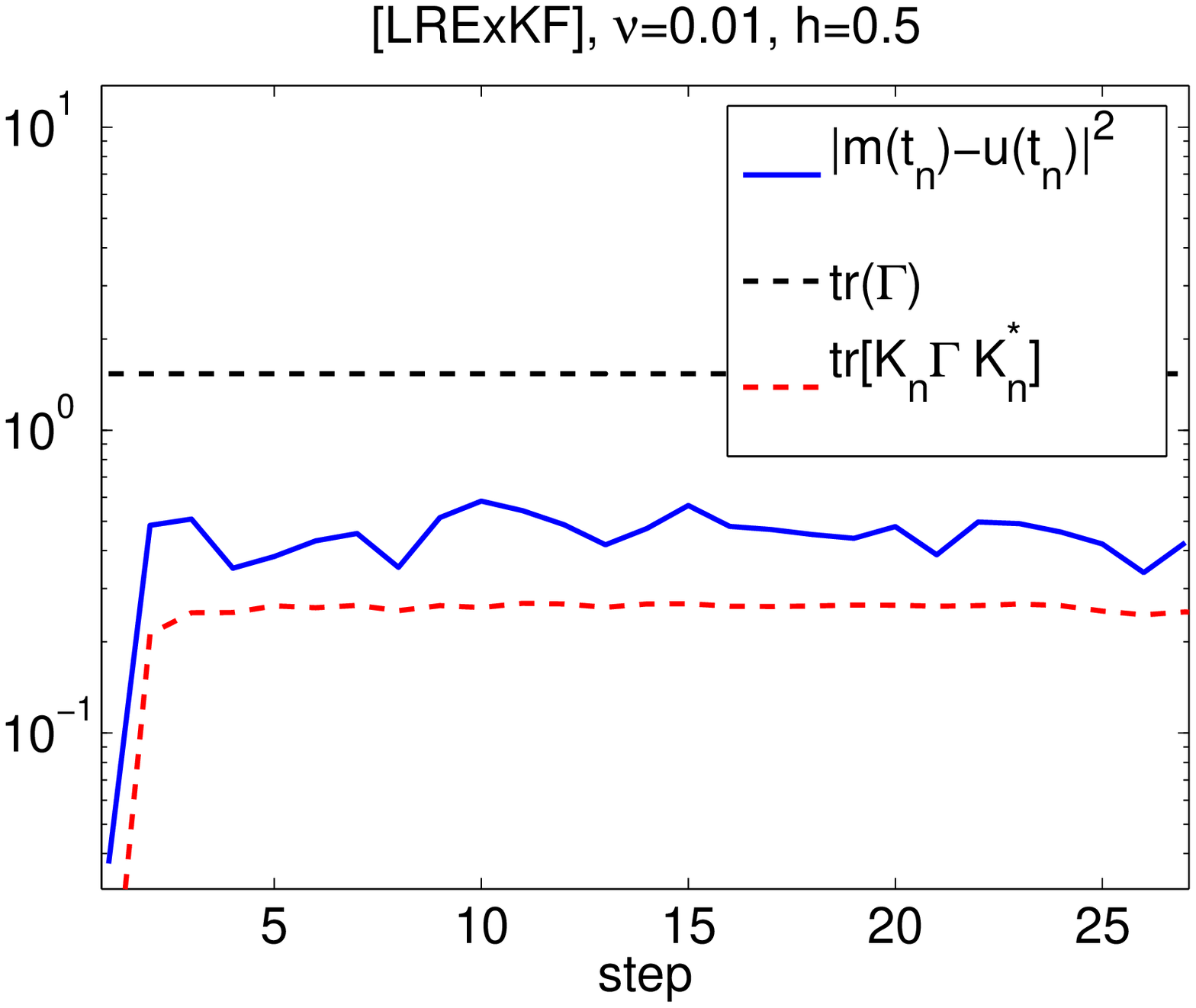}
\includegraphics[width=.45\textwidth]{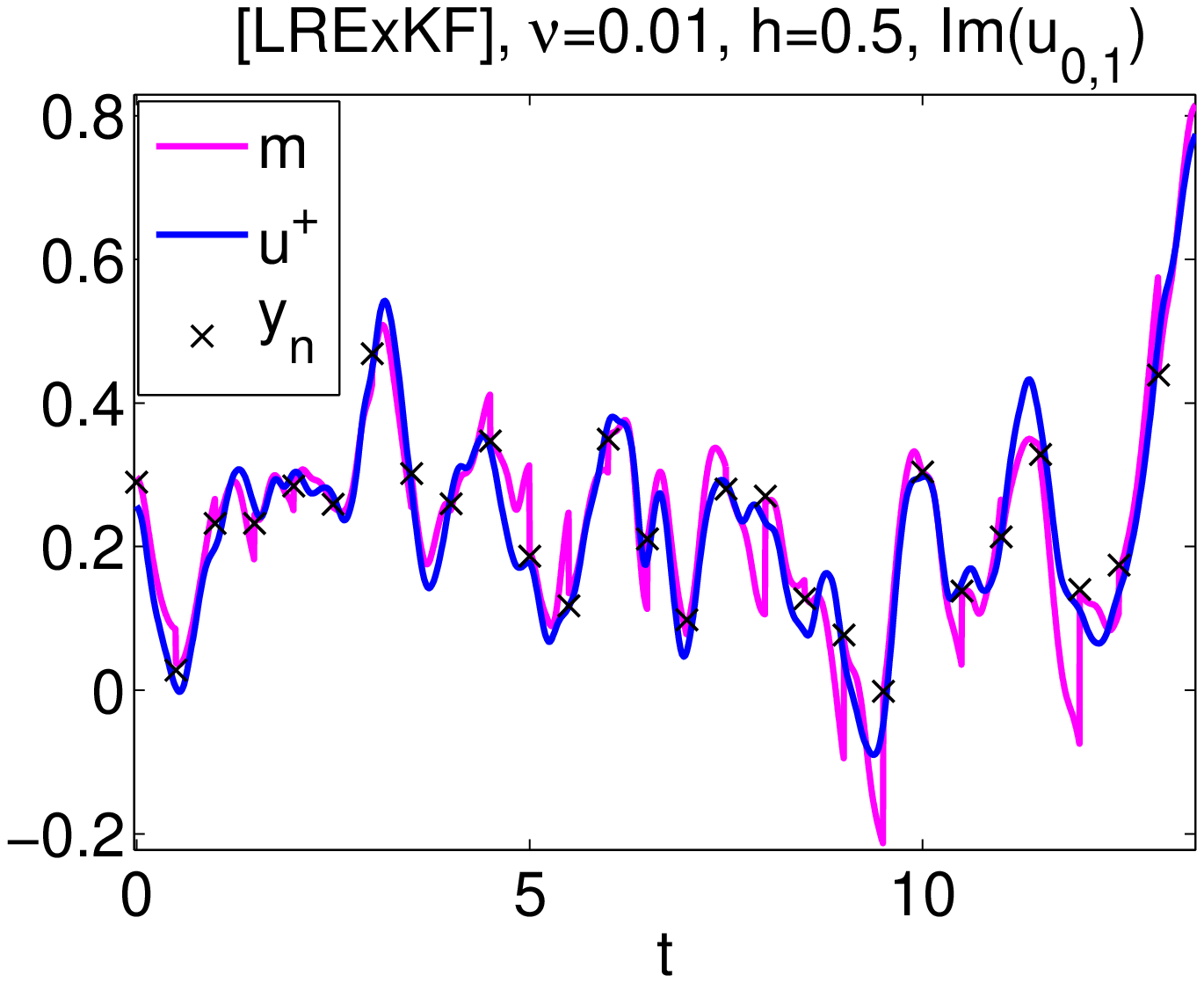}
\includegraphics[width=.45\textwidth]{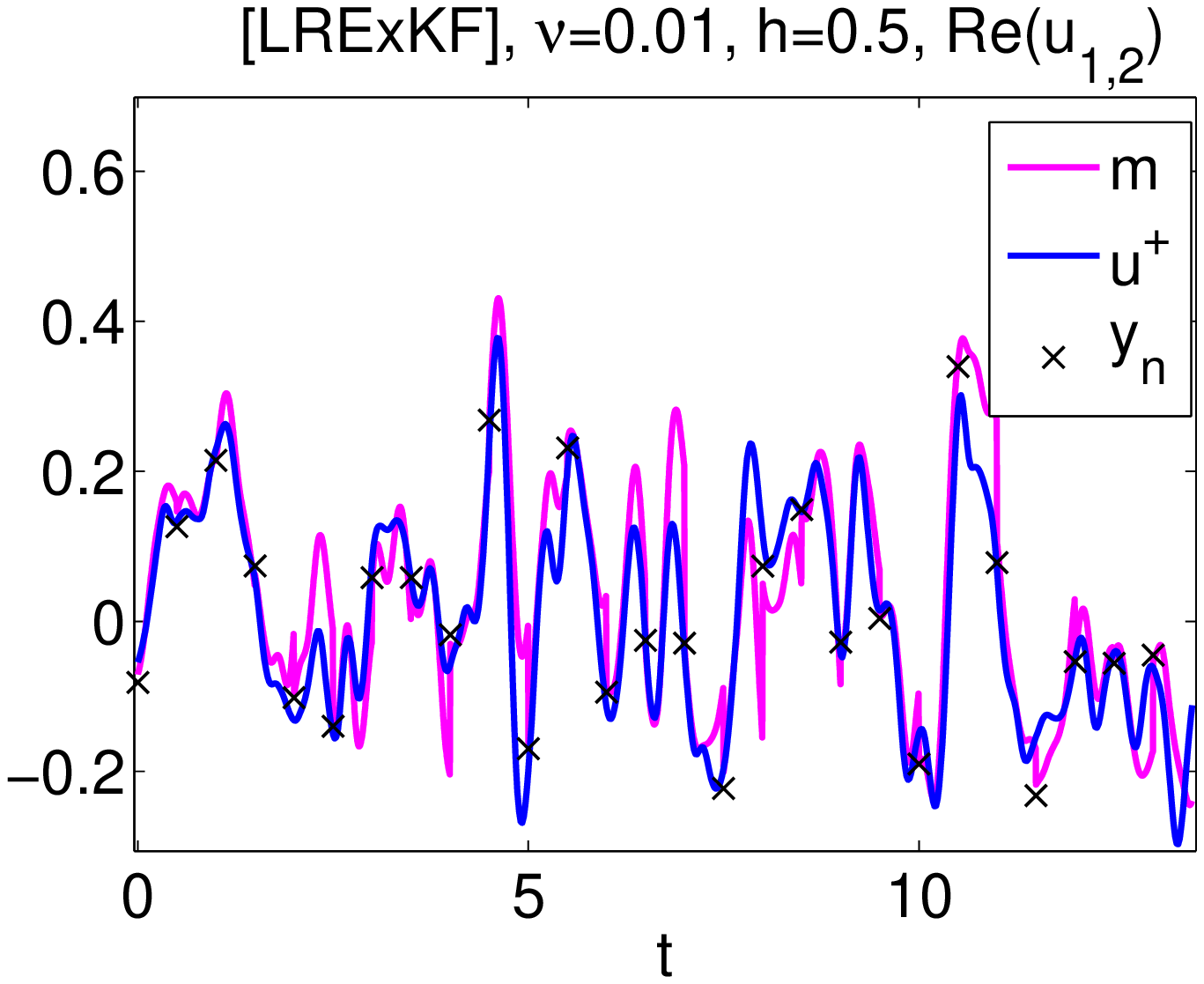}
\includegraphics[width=.45\textwidth]{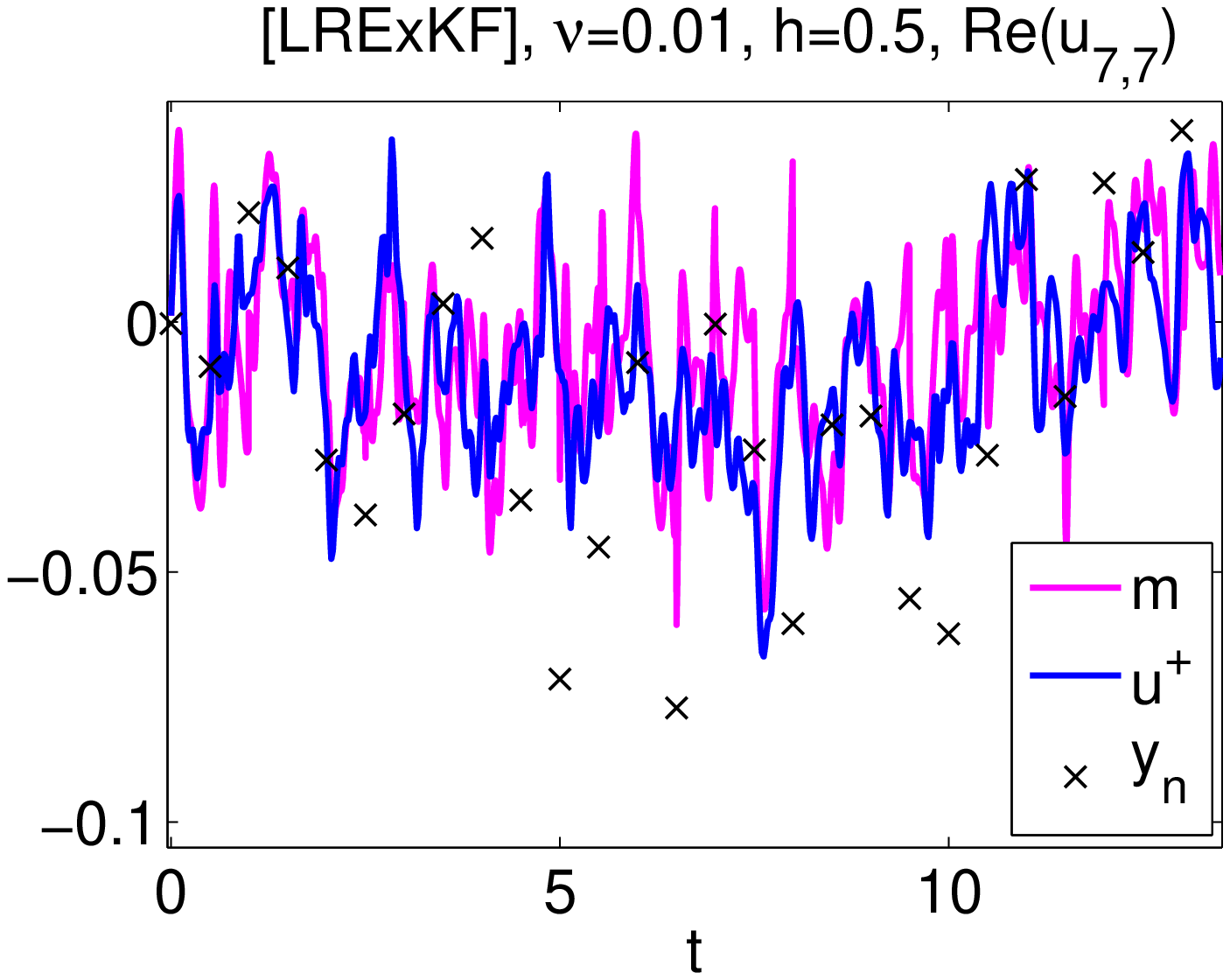}
\caption{Example of a variance-inflated stablilized trajectory 
(updated with model b from Section \ref{fdf} on the complement of the
low-rank approximation) 
for [LRExKF] with the same external parameters as in
  Fig. \ref{fdf_unstable}.
Panels are the same as in Fig. \ref{fdf_unstable}.}
\label{fdf_stable}
\end{figure*}

Examples of an unstable instance of 3DVAR
and the corresponding stabilized filter, [3DVAR], 
are depicted in Figures \ref{var_unstable} 
and \ref{var_stable}, respectively, with $\nu=0.01, h=0.2.$ 
In this regime the dynamics are strongly chaotic.
The first point to note is that both simulations give
rise to an error which exceeds the lower bound \eqref{eq:lower};
and that the unstable algorithm exceeds the desired bound
\eqref{eq:upper}, whilst the stabilized algorithm does not;
note also that the stabilized algorithm output is plotted
over a longer time-interval than the original algorithm.
A second noteworthy point relates to the power of using the
dynamical model: this is manifest in the bottom right panels
of each figure, in which the trajectory of a high wavenumber
mode, close to the forcing frequency, is shown.
The assimilation performs remarkably well for the trajectory of 
this wavenumber relative to the observations 
in the stabilized case, owing to the high weight on 
the dynamics and stability of the
dynamical model for that wavenumber. 
Examples of an unstable instance of LRExKF
and the corresponding stabilized filter, [LRExKF], 
are depicted in Figures \ref{fdf_unstable} 
and \ref{fdf_stable}, respectively, with $\nu=0.01, h=0.5$. 
The behaviour illustrated is very similar to that exhibited
for 3DVAR and [3DVAR].

In the following tables we 
make a comparison between the original form of the filters 
and their stabilized forms,
using the gold standard
Bayesian posterior distribution as the desired outcome.
Table \ref{turbh1s} 
shows data for $h=0.02$ and $0.2$ 
with $T=0.2$ fixed.  
Tables \ref{turbh10_6s} and \ref{turbh25_3s} 
show data for $h=0.2$ and $0.5$ with $T=1$ fixed.
We focus our discussion on the approximation of the mean.
It is noteworthy that, on the shorter time horizon $T=0.2$,
the stabilized algorithms are less accurate with respect to 
the mean than their
original counterparts, for both values of observation time $h$;
this reflects a lack of accuracy caused by inflating
the variance.
As would be expected, however,
this behaviour is reversed on longer time-intervals, as
is shown when $T=1.0$, relfecting enhanced stability cased
by inflating the variance.
Table \ref{turbh10_6s} shows the case $T=1.0$ with $h=0.2$, 
and the stabilized version of
3DVAR outperforms the original version,
although the stabilized versions of EnKF and LRExKF are not 
as accurate as the original version.  In Table \ref{turbh25_3s},
with $h=0.5$ and $T=1.0$, the stabilized 
versions improve upon the
original algorithms in all three cases.
Furthermore, in Table \ref{turbh25_3s}, we also display
the FDF showing that, without any stabilization, this
outperforms the other three filters and their stabilized
counterparts.

\section{Conclusion}
\label{conclusion}

Incorporating noisy data into uncertain computational
models presents a major challenge in many areas of the
physical sciences, and in atmospheric modelling and
NWP in particular. Data assimilation algorithms in  
NWP have had measurable positive impact on forecast skill.
Nonetheless, assessing the ability of these algorithms
to forecast {\em uncertainty} is more subtle. It is
important to do so, however, especially as prediction is
pushed to the limits of its validity in terms of
time horizons considered, or physical processes
modelled. In this paper we have proposed an approach to 
the evaluation of the ability of data assimilation
algorithms to predict uncertainty. 
The cornerstone of our approach is to adopt a fully
non-Gaussian Bayesian perspective in which the probability
distribution of the system state over a time horizon,
given data over that time horizon, plays a pivotal role:
we contend that algorithms should be evaluated by their
ability to reproduce this probability distribution, or
important aspects of it, accurately.

In order to make this perspective useful it is necessary to find
a model problem which admits complex behaviour reminiscent
of atmospheric dynamics,
whilst being sufficiently small to allow computation of
the Bayesian posterior distribution, so that data assimilation
algorithms can be compared against it. 
Although MCMC sampling of the posterior can, 
in principle, recover any distribution, 
it becomes prohibitively expensive for multi-modal 
distributions, depending on the 
energy barriers between modes. However for unimodal
problems, state-of-the-art sampling techniques
allow fully resolved MCMC computations to be undertaken.
We have found that the 2D Navier-Stokes equations 
provide a model for which the posterior distribution
may be accurately sampled using MCMC,
in regimes where the dynamics is 
stationary and where it is strongly chaotic.
We have confined our attention to strong constraint models,
and implemented a range of variational and filtering methods,
evaluating them by their ability to reproduce the Bayesian
posterior distribution. The set-up is such that the
Bayesian posterior is unimodal and approximately Gaussian.
Thus the evaluation is undertaken by comparing the mean
and covariance structure of the data assimilation algorithms
against the actual Bayesian posterior mean and covariance.
Similar studies were undertaken in the context of a subsurface
geophysical inverse problem in \cite{Liu_Oliver_2003},
although the conclusions were less definitive. It would
be interesting to revisit such subsurface geophysical 
inverse problems
using the state-of-the-art MCMC techniques adopted here,
in order to compute the posterior distribution. 
Moreover it would be interesting to
conduct a study, similar to that undertaken here, for
models of atmospheric dynamics such as Lorenz-96, or
a quasi-geostrophic models, which admit baroclinic instabilities.

These studies, {\it under the assumption of a well-defined posterior
  probability distribution}, lead to four conclusions: (i) most filtering
and variational algorithms do a reasonably good job of
reproducing the mean; (ii) for most of the filtering and
variational algorithms studied and implemented here 
there are circumstances
where the approximations underlying the ad hoc filters are
not justified and they then fail to reproduce covariance 
information with any accuracy
(iii) most filtering algorithms exhibit
instability on longer time-intervals causing them to lose
accuracy in even mean prediction; (iv) filter stabilization,
via variance inflation of one sort or the other,
ameliorates this instability and can improve long-term accuracy
of the filters in predicting the mean,
but can reduce the accuracy on short time intervals
and of course makes it impossible to predict the 
covariance.
In summary most data assimilation algorithms used in practice 
should be viewed with caution when using them to make claims 
concerning uncertainty although, when
properly tuned, they will frequently
track the signal mean accurately for fairly long
time intervals.  These conclusions are intrinsic to the
algorithms, and result from the nature of the
approximations made in order to create tractable online
algorithms; the basic conclusions 
are not expected to change
by use of different dynamical models, or by modifying the
parameters of those algorithms. 

\begin{figure*}
\includegraphics[width=0.5\textwidth]{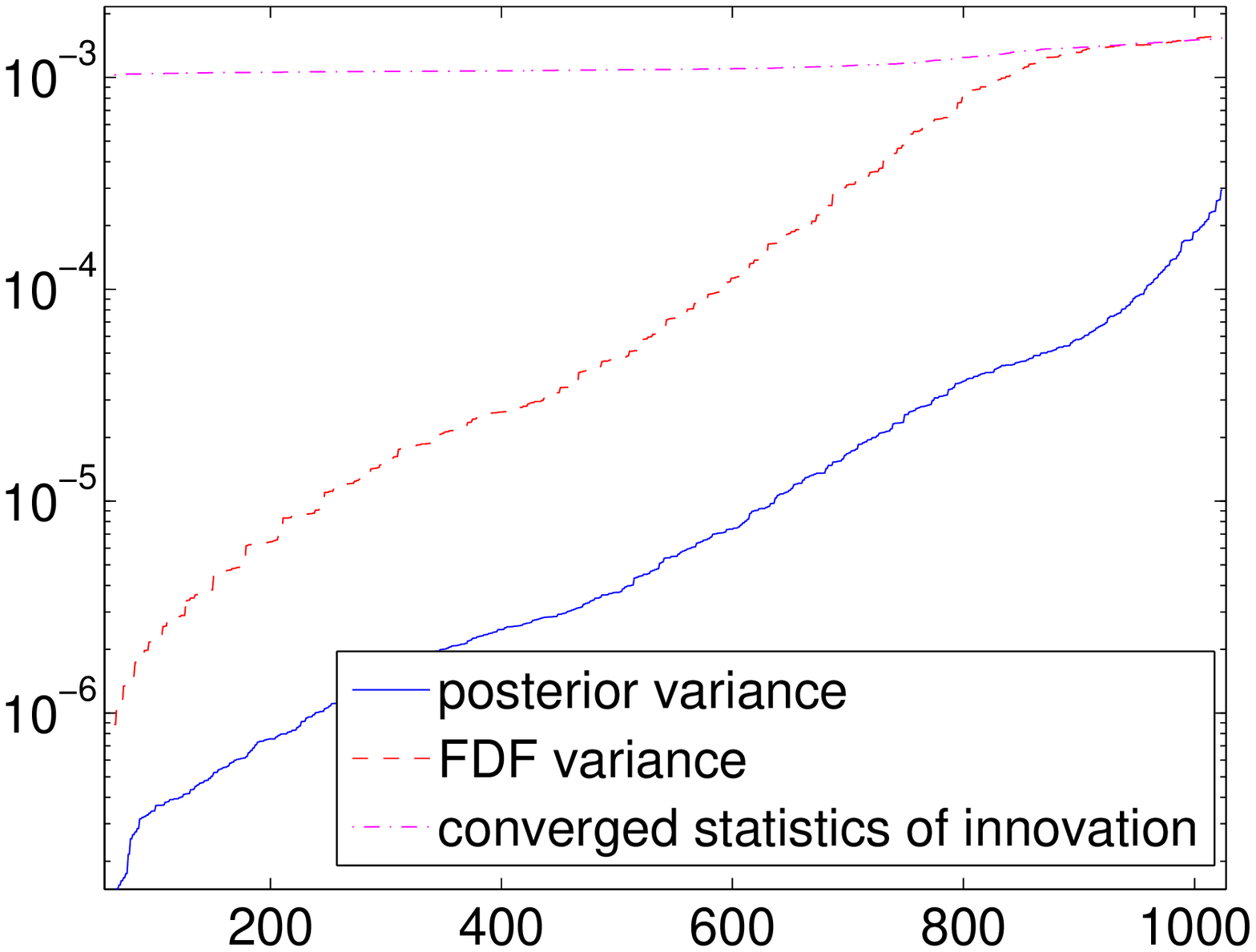}
\includegraphics[width=0.5\textwidth]{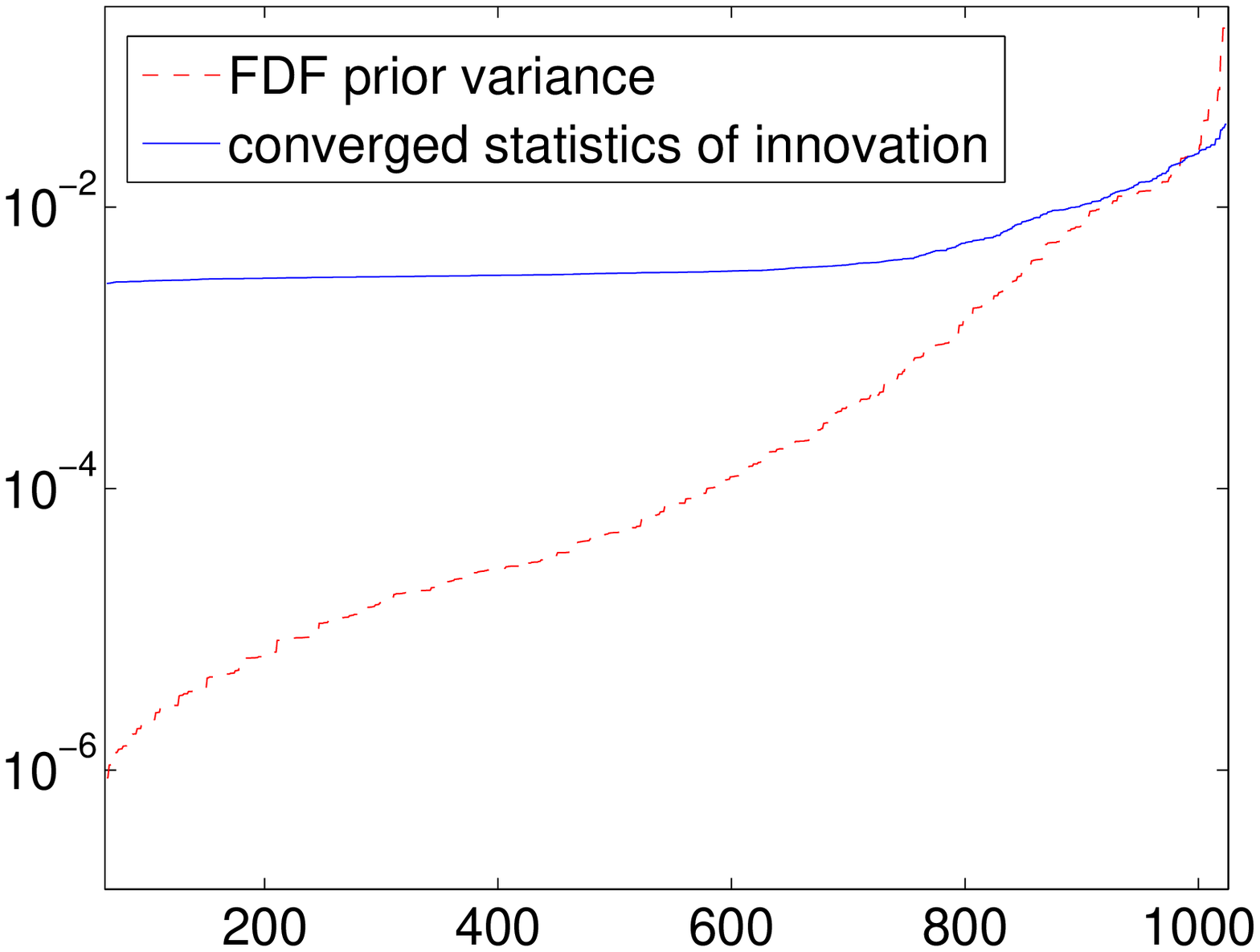}
\caption{The left and right panels, repsectively, show the 
posterior and prior of the covariance from converged innovation
 statistics from the cycled 3DVAR algorithm,
in comparison to the converged covariance from the FDF algorithm,
and the posterior distribution. }
\label{rankhistos}
\end{figure*}

Finally we note that we have not addressed in this paper
the important but complicated issue of how to choose the
prior distribution on the initial condition. We finish
with some remarks concerning this.
The ``accuracy of the spread'' of the prior is often
monitored in practice with a rank histogram \cite{anderson1996method}.
This can be computed even in the absence of an ensemble for any method
in the class of those discussed here, by partitioning the real line in
bins according to the assumed Gaussian prior density.  It is important
to note that uniform component-wise rank histograms in each direction 
guarantee that there are no
directions in which the variance is consistently underestimated, and
this should therefore be sufficient for stability.  It is also
necessary for the accurate approximation of the Bayesian posterior
distribution, but by no means sufficient 
\cite{hamill2000comparison}.  Indeed, one can 
iteratively compute a constant prior with the cycled 3DVAR algorithm
\cite{hamill2000comparison} such that the estimator from the algorithm 
will have statistics consistent with the constant prior used in the 
algorithm.  The estimator produced by this algorithm is guaranteed by 
construction to yield uniform rank histograms of the type described
above, and yet the actual prior coming from the posterior 
at the previous time is not constant, so this cannot be a
good approximation of the actual prior.  
See Fig. \ref{rankhistos} for an image of the 
variance which is consistent with the statistics of the estimator 
over $100$ iterations of 3DVAR with $\nu=0.01$ and $h=0.5$, 
as compared with the 
prior, posterior, and converged FDF variance at $T=1$.  Notice FDF
overestimates in the high-variance directions, and underestimates in 
the low-variance directions (which correspond in our case to the 
unstable and stable directions, respectively).  The RMSE of 3DVAR with
constant converged FDF variance is smaller than with constant variance from
converged statistics, and yet the former clearly will yield component-wise rank
histograms which appear to always underestimate the ``spread'' in the  
low-variance, stable directions, and overestimate in the
high-variance, unstable directions.  It is also noteworthy that the
FDF variance accurately recovers the decay of the posterior variance, but
is about an order of magnitude larger.
Further investigation of how to initialize
statistical forecasting algorithms clearly remains a
subject presenting many conceptual and practical challenges.

\begin{table*}
\begin{center}
\begin{tabular}{|c||c|c|c|c|c|} \hline
$h=0.02$        &$e_{mean}$    &$e_{variance}$&$e_{truth}$   &$e_{obs}$     &$e_{map}$      \\ \hline \hline
 MCMC($t=0$)  & 0 & 0 & 0.0331468 & 0.337233 & 0.000731645 \\ \hline
 4DVAR($t=0$) & 0.000731491 & 0.0932748 & 0.0331531 & 0.310411 & 0 \\ \hline
 MCMC($t=T$)  & 0 & 0 & 0.0423943 & 0.32224 & 0.00130105 \\ \hline
 4DVAR($t=T$) & 0.00130112 & 0.045048 & 0.042431 & 0.322306 & 0 \\ \hline
3DVAR        & 0.0634553 & 6.34057 & 0.063289 & 0.321959 & 0.0634026 \\ \hline
 FDF       & 0.165732 & 28.9155 & 0.175397 & 0.307159 & 0.165844 \\ \hline
LRExKF       & 0.00599214 & 0.030054 & 0.0416529 & 0.322277 & 0.0054415 \\ \hline
EnKF       & 0.035271 & 0.274428 & 0.0523566 & 0.323074 & 0.0354624 \\ \hline
truth ($t=0$)& 0.0331468 & - & 0 & 0.335933 & 0.0361395 \\ \hline
 truth ($t=T$)& 0.0423943 & - & 0 & 0.339539 & 0.0429021 \\ \hline
\hline \hline
$h=0.1$        &$e_{mean}$    &$e_{variance}$&$e_{truth}$   &$e_{obs}$     &$e_{map}$      \\ \hline \hline
 MCMC($t=0$)  & 0 & 0 & 0.0496982 & 0.294743 & 0.000815864 \\ \hline
 4DVAR($t=0$) & 0.000815762 & 0.0287498 & 0.0497009 & 0.280425 & 0 \\ \hline
 MCMC($t=T$)  & 0 & 0 & 0.0698665 & 0.35798 & 0.00306996 \\ \hline
 4DVAR($t=T$) & 0.00307105 & 0.0118785 & 0.06983 & 0.358094 & 0 \\ \hline
3DVAR        & 0.159393 & 2.2339 & 0.203165 & 0.374188 & 0.159658 \\ \hline
 FDF          & 0.200044 & 13.259 & 0.215136 & 0.308921 & 0.200045 \\ \hline
LRExKF       & 0.023073 & 0.0313686 & 0.0766505 & 0.357915 & 0.0215118 \\ \hline
EnKF        & 0.0539001 & 0.174878 & 0.109402 & 0.358301 & 0.0543726 \\ \hline
truth ($t=0$)& 0.0496982 & - & 0 & 0.303742 & 0.0541391 \\ \hline
 truth ($t=T$)& 0.0698665 & - & 0 & 0.368335 & 0.0705546 \\\hline
\hline \hline
$h=0.2$        &$e_{mean}$    &$e_{variance}$&$e_{truth}$   &$e_{obs}$     &$e_{map}$      \\ \hline \hline
 MCMC($t=0$)  & 0 & 0 & 0.0459125 & 0.293686 & 0.00122936 \\ \hline
  4DVAR($t=0$) & 0.00183617 & 0.0231955 & 0.0462013 & 0.281137 & 0 \\ \hline
MCMC($t=T$)  & 0 & 0 & 0.072738 & 0.352456 & 0.00385795 \\ \hline
 4DVAR($t=T$) & 0.00386162 & 0.0196227 & 0.0723178 & 0.352145 & 0 \\ \hline
3DVAR        & 0.285461 & 1.72154 & 0.300853 & 0.38443 & 0.286161 \\ \hline
 FDF       & 0.202274 & 10.7793 & 0.203287 & 0.316707 & 0.202862 \\ \hline
LRExKF       & 0.0750908 & 0.0547417 & 0.0886932 & 0.35073 & 0.0726792 \\ \hline
EnKF        & 0.0964053 & 0.0948967 & 0.113806 & 0.352625 & 0.0962341 \\ \hline
 truth ($t=0$)& 0.0459125 & - & 0 & 0.301899 & 0.0496251 \\ \hline
 truth ($t=T$)& 0.072738 & - & 0 & 0.368331 & 0.0720492 \\ \hline
\end{tabular} 
\end{center}
\caption{Same as Table \ref{stationaryh1}, except for strongly chaotic
  regime with $\nu=0.01$, $T=0.2$, and $h=0.02$ (top), 
$0.1$ (middle) and $0.2$ (bottom).}
\label{turbh1}
\end{table*}

\begin{table*}
\begin{center}
\begin{tabular}{|c||c|c|c|c|c|} \hline
$h=0.2$        &$e_{mean}$    &$e_{variance}$&$e_{truth}$   &$e_{obs}$     &$e_{map}$      \\ \hline \hline
 MCMC($t=0$)  & 0 & 0 & 0.0322397 & 0.294722 & 0.00122667 \\ \hline
 4DVAR($t=0$) & 0.00122657 & - & 0.0316494 & 0.280742 & 0 \\ \hline
 MCMC($t=T$)  & 0 & 0 & 0.0480924 & 0.27997 & 0.00484999 \\ \hline
 4DVAR($t=T$) & 0.0048519 & - & 0.0474821 & 0.279995 & 0 \\ \hline
 3DVAR        & 0.35571 & 3.17803 & 0.357351 & 0.419614 & 0.35557 \\ \hline
FDF  & 0.141426 & 19.2983 & 0.152064 & 0.260197 & 0.142169 \\ \hline
LRExKF       & 0.101179 & 0.28308 & 0.0900697 & 0.291704 & 0.101287 \\ \hline
EnKF         & 0.202724 & 0.230518 & 0.173947 & 0.320302 & 0.202665 \\ \hline
truth ($t=0$)& 0.0322397 & - & 0 & 0.303376 & 0.0272922 \\ \hline
 truth ($t=T$)& 0.0480924 & - & 0 & 0.281553 & 0.0474964 \\ \hline
\hline \hline
$h=0.5$        &$e_{mean}$    &$e_{variance}$&$e_{truth}$   &$e_{obs}$     &$e_{map}$      \\ \hline \hline
 MCMC($t=0$)  & 0 & 0 & 0.0318531 & 0.293871 & 0.0030989 \\ \hline
 4DVAR($t=0$) & 0.00309769 & - & 0.0313382 & 0.280152 & 0 \\ \hline
 MCMC($t=T$)  & 0 & 0 & 0.0460821 & 0.288812 & 0.00831516 \\ \hline
 4DVAR($t=T$) & 0.00831886 & - & 0.0448424 & 0.289043 & 0 \\ \hline
 3DVAR        & 0.458527 & 1.8214 & 0.45353 & 0.487658 & 0.460144 \\ \hline
 FDF          & 0.189832 & 11.4573 & 0.19999 & 0.25111 & 0.191364 \\ \hline
  LRExKF       & 0.644427 & 0.325391 & 0.650004 & 1.22145 & 0.646233 \\ \hline
EnKF       & 0.901703 & 0.554611 & 0.895878 & 0.908817 & 0.902438 \\ \hline
truth ($t=0$)& 0.0318531 & - & 0 & 0.303185 & 0.0269929 \\ \hline
 truth ($t=T$)& 0.0460821 & - & 0 & 0.294524 & 0.0448046 \\ \hline
\end{tabular} 
\end{center}
\caption{Same as Table \ref{turbh1}, except $T=1$, and $h=0.2$
  (top) and $h=0.5$ (bottom).  
The variance is ommitted from the 4DVAR solutions here, because we are
  unable to attain solution with zero derivative. We must
    note here that we have taken the approach of differentiating and
    then discretizing.  Therefore, over longer time intervals such as
    this, the error between the discretization of the analytical derivative and 
derivative of the finite-dimensional discretized forward map 
accumulates and the derivative of the objective function is no longer 
well-defined because of this error.  Nonetheless, we confirm that 
we do obtain the MAP estimator because the MCMC run does
not yield any point of higher probability.}
\label{turbh10_6}
\end{table*}

\begin{acknowledgment} 
Both authors are grateful to the referees for numerous
suggestions which have improved the presentation of
this material; In particular, we thank Chris Snyder.
KJHL is grateful to the EPSRC for funding. AMS
is grateful to EPSRC, ERC and ONR for funding.
\end{acknowledgment}

\section{Appendix: Some numerical details}
\label{numerics}

Here we provide some details of the numerical
algorithms underlying the computations 
which we present in the main body of the paper.
First, we will describe the
numerical methods used for
the dynamical model. Secondly we study the adjoint solver.
Thirdly we discuss various issues related to the resulting
optimization problems and large linear systems encountered.
Finally we discuss the
MCMC method used to compute the gold standard posterior
probability distribution. 

In the {\em dynamical and observational models} the 
forcing in Eq. \ref{navier_stokes} is taken to be $f=\nabla^{\perp}\psi$,
where $\psi=\cos(k \cdot x)$ and $\nabla^{\perp}=J\nabla$ with $J$
the canonical skew-symmetric matrix, and $k=(1,1)$ for stationary ($\nu=0.1$)
regime, while $k=(5,5)$ for the strongly chaotic regime in order to allow
an upscale cascade of energy.  
Furthermore, we set the observational noise to white
noise $\Gamma= \gamma^2 I$, where $\gamma = 0.04$ is chosen as 
10\% of the maximum standard deviation of the strongly chaotic 
dynamics, and we choose 
an initial smoothness prior $\cC_0 = A^{-2}$,
where $A$ is the Stokes operator.  
We notice that only the observations on the 
unstable manifold of the underlying solution map
need to be assimilated.  A similar observation
was made in \cite{chorin2004dimensional} 
in the context of particle filters.
Our choice of prior and observational covariance reflect this
in the sense that the ratio of the prior to the observational
covariance is larger for smaller wavenumbers (and greater than 
1, in particular), in which the unstable manifold has support, 
while this ratio tends to zero as $|k| \rightarrow \infty$.
The initial mean,
or background state, is chosen as $m_0 \sim \cN(u^\dagger,\cC_0)$,
where $u^\dagger$ is the true initial condition.
In the case of strongly chaotic
dynamics it is taken as an arbitrary point on the attractor obtained by
simulating an arbitrary initial condition until statistical
equilibrium.  The initial condition for the case of stationary
dynamics is taken as a draw from the Gaussian prior, 
since the statistical equilibrium is the trivial one.

Our numerical method for the dynamical model
is based on a Galerkin approximation of
the velocity field in a divergence-free Fourier basis. 
We use a modification of a
fourth-order Runge-Kutta method,  ETD4RK \cite{cox2002exponential}, 
in which the heat semi-group 
is used together with Duhamel's principle to solve exactly
for the diffusion term.  
A spectral Galerkin method \cite{hesthaven2007spectral} is used in which the 
convolutions arising from products in the nonlinear term are computed 
via FFTs.
We use a double-sized domain in each dimension, buffered with
zeros, resulting in $64^2$ grid-point FFTs, and only half the modes
are retained when transforming back into spectral space in 
order to prevent de-aliasing which is avoided
as long as fewer than 2/3 the modes are retained.
Data assimilation
in practice always contends with poor spatial resolution, particularly
in the case of the atmosphere in which there are many billions of 
degrees of freedom.  For us the important resolution
consideration is that the unstable modes, 
which usually have long spatial scales and 
support in low wave-numbers, are resolved.
Therefore, our objective here is not to obtain high spatial-resolution
but rather to obtain high temporal-resolution in the sense
of reproducibility.  We would like the divergence of two close-by
trajectories to be dictated by instability in the dynamical model rather
than the numerical time-stepping scheme.  

It is also important that we have
accurate {\em adjoint solvers}, and this is 
strongly linked to the accuracy of
the forward solver.  
The same time-stepper is used to solve the adjoint
equation, with twice the time-step of the forward solve, since the 
forward solution is required at half-steps in order to implement this
method for the non-autonomous adjoint solve.  
Many issues can arise in the
implementation of adjoint, or costate methods 
\cite{banks1992computational,vogel1995analysis} 
and the practitioner should be aware of these.  The easiest way to 
ensure convergence is to test that the tangent linearized map 
is indeed the linearization of the solution map and then confirm that 
the adjoint is the adjoint to a suitable threshold.
We have taken the approach of ``optimize then
  discretize'' here, and as such our adjoint model is the
  discretization of the analytical adjoint.  This effect becomes
  apparent in the accuracy of the linearization for longer time 
intervals, and we are no longer able to compute accurate gradients 
and Hessians as a result.

Regarding {\em linear algebra and optimization}
issues we make the following observations.
A Krylov method (GMRES) is used for linear
solves in the Newton method for 4DVAR, and the Arnoldi method is used
for low-rank covariance approximations in LRExKF and for the filtering
time $T$ covariance approximation in 4DVAR.  
The LRExKF always sufficiently captures more than $99\%$ of the full rank 
version as measured in Frobenius (matrix $l^2$) norm. The initial Hessian in 
4DVAR and well as the ones occuring within Newton's method
are computed by finite difference.  Using a gradient flow 
(preconditioned steepest descent)
computation, we obtain an approximate minimizer close to the
actual minimizer and then a preconditioned 
Newton-Krylov nonlinear fixed-point solver is used ({\sc nsoli}
\cite{kelley2003solving}).  
This approach is akin to the Levenburgh-Marquardt algorithm.
See \cite{trefethen1997numerical}
and \cite{saad1996iterative} for overviews of the linear algebra
and \cite{nocedal1999numerical} for an overview of optimization.
Strong constraint 4DVAR can be computationally challenging
and, although we do not do so here, 
it would be interesting to study weak constraint 4DVAR
from a related perspective;  see \cite{brocker2010variational} 
for a discussion of weak constraint 4DVAR in continuous time.
It is useful to employ benchmarks in order to confirm gradients
are being computed properly when implementing optimizers,
see for example \cite{lawless2003comparison}.

Finally, we comment on the {\em MCMC} computations
which, of all the algorithms implemented here, lead 
to the highest computational cost. This, of course, is
because it fully resolves the posterior distribution of
interest whereas the other algorithms use crude
approximations, the consequences of which we 
study by comparison with accurate MCMC results. 
Each time-step requires $4$ function evaluations, and each function
evaluation requires $8$ FFTs, so it costs $32$ FFTs for each
time-step.  We fix the lengths of paths at $40$ time-steps for most
of the computations, but nonetheless, this is on the order of
$1000$ FFTs per evaluation of the dynamical model.  
If a $64^2$ FFT takes $1$ ms, then this amounts 
to $1$ s per sample.  Clearly this is a hurdle as it would take on the
order of $10$ days to obtain on the order of millions of samples in serial.  
We overcome this by using the MAP estimator (4DVAR solution)
as the initial condition 
in order to accellerate burn-in, and then run independent batches of 
$10^4$ samples in parallel with 
independent seeds in the random number generator.
We also minimize computional effort within the method by
employing the technique of early rejection 
introduced by \cite{haario_earlyrejection} which
means that rejection can be detected before
the forward computation required for evaluation
of $\Phi$ reaches the end of the assimilation time window;
the computation can then be stopped 
and hence computational savings made.

It is important to recognize that
we cannot rely too heavily on results of MCMC with smaller
relative norm than $10^{-3}$ for the mean or $10^{-2}$ for the variance,
because we are bound to $\mathcal{O}(N^{-1/2})$ 
convergence and it is already prohibitively expensive to get several
million samples.  More than $10^7$ is not tractable.  Convergence is
measured by a version of MSPRF \cite{brooks1998general},
$ev_{1:8}=||{\rm var}[u_1(t)]-{\rm var}[u_8(t)]||/||{\rm var}[u_1(t)]||$, where
$u_1$ corresponds to sample statistics with $1$ chain and 
$u_8$ corresponds to sample statistics over $8$ chains.
We find $ev_{1:8}=\mathcal{O}(10^{-2})$ for
$N=3.2 \times 10^5$ samples in each chain.
If we define
$em_{1:8}=||\bbE[u_1(t)]-\bbE[u_8(t)]||/||\bbE[u_1(t)]||$,
then we have $em_{1:8}=\mathcal{O}(10^{-3})$.

\begin{table*}
\begin{center}
\begin{tabular}{|c||c|c|c|c|c|} \hline
h=0.02        &$e_{mean}$    &$e_{variance}$&$e_{truth}$   &$e_{obs}$     &$e_{map}$      \\ \hline \hline
 3DVAR        & 0.0634553 & 6.34057 & 0.063289 & 0.321959 & 0.0634026 \\ \hline
 [3DVAR]      & 0.142759 & 22.2668 & 0.153141 & 0.309838 & 0.143005 \\ \hline
EnKF         & 0.035271 & 0.274428 & 0.0523566 & 0.323074 & 0.0354624 \\ \hline
[EnKF]       & 0.167776 & 28.1196 & 0.175359 & 0.304352 & 0.167919 \\ \hline
h=0.2        &$e_{mean}$    &$e_{variance}$&$e_{truth}$   &$e_{obs}$     &$e_{map}$      \\ \hline \hline
3DVAR        & 0.285461 & 1.72154 & 0.300853 & 0.38443 & 0.286161 \\ \hline
[3DVAR]      & 0.195222 & 6.33608 & 0.204883 & 0.339108 & 0.196339 \\ \hline
LRExKF       & 0.0750908 & 0.0547417 & 0.0886932 & 0.35073 & 0.0726792 \\ \hline
[LRExKF]     & 0.156973 & 7.64123 & 0.169354 & 0.310298 & 0.156596 \\ \hline
EnKF         & 0.137844 & 0.372259 & 0.159744 & 0.353934 & 0.137969 \\ \hline
[EnKF]      & 0.248081 & 6.34903 & 0.267746 & 0.368067 & 0.249475 \\ \hline
\end{tabular} 
\end{center}
\caption{The data of unstable algorithms from Table \ref{turbh1}
  ($\nu=0.01$, $T=0.2$) 
are reproduced above (with $h=0.02$(top) and $h=0.2$(bottom)), 
along with the respective stabilized versions in brackets.  
Here the stabilized versions usually perform worse.  Note that 
over longer time scales, the unstabilized version will diverge 
from the truth, while the stabilized one remains close.}
\label{turbh1s}
\end{table*}

\begin{table*}
\begin{center}
\begin{tabular}{|c||c|c|c|c|c|} \hline
h=0.2        &$e_{mean}$    &$e_{variance}$&$e_{truth}$   &$e_{obs}$     &$e_{map}$      \\ \hline \hline
3DVAR        & 0.35571 & 3.17803 & 0.357351 & 0.419614 & 0.35557 \\ \hline
[3DVAR]      & 0.131964 & 11.5997 &0.135572 & 0.277895 & 0.133265 \\ \hline
LRExKF       & 0.101179 & 0.28308 & 0.0900697 & 0.291704 & 0.101287 \\ \hline
[LRExKF]     & 0.12962 & 16.3692 & 0.13592 & 0.256617 & 0.129742 \\ \hline
EnKF         & 0.0736613 & 0.276947 & 0.0755232 & 0.282247 & 0.0742144 \\ \hline
[EnKF]      & 0.1231 & 14.8557 & 0.133171 & 0.261061 & 0.124203 \\ \hline
\end{tabular} 
\end{center}
\caption{Same as Table \ref{turbh1s}, except $T=5h=1$ 
and $h=0.2$.
[3DVAR] performs better with respect to the mean.}
\label{turbh10_6s}
\end{table*}

\begin{table*}
\begin{center}
\begin{tabular}{|c||c|c|c|c|c|} \hline
h=0.5        &$e_{mean}$    &$e_{variance}$&$e_{truth}$   &$e_{obs}$     &$e_{map}$      \\ \hline \hline
3DVAR        & 0.458527 & 1.8214 & 0.45353 & 0.487658 & 0.460144 \\ \hline
 [3DVAR]      & 0.27185 & 6.62328 & 0.285351 & 0.307263 & 0.274663 \\ \hline
 LRExKF       & 0.644427 & 0.325391 & 0.650004 & 1.22145 & 0.646233 \\ \hline
[LRExKF]     & 0.201327 & 11.2449 & 0.207526 & 0.244101 & 0.201081 \\ \hline
EnKF        & 0.901703 & 0.554611 & 0.895878 & 0.908817 & 0.902438 \\ \hline
[EnKF]      & 0.169262 & 4.07238 & 0.17874 & 0.244571 & 0.170245 \\ \hline
 FDF          & 0.189832 & 11.4573 & 0.19999 & 0.25111 & 0.191364 \\ \hline
\end{tabular} 
\end{center}
\caption{Same as Table \ref{turbh10_6s}, except $h=0.5$.  
All stabilized algorithms now perform better with respect to the mean.
[LRExKF] above uses 50 eigenvectors in the low rank representation, 
and performs {\it worse} for larger number, indicating that the
improvement is due largely to the FDF component.
The stable FDF data are included here as well, since FDF is now
competitive as the optimal
  algorithm in terms of mean estimator.  This is expected to persist
  for larger time windows and lower frequency observations, since the
  LRExKF is outside of the regime of validity, as shown in Figure \ref{spectrum}.}
\label{turbh25_3s}
\end{table*}

 \bibliographystyle{ametsoc}

\bibliography{posteriorbib}

\end{document}